\documentclass[journal]{IEEEtran}
\IEEEoverridecommandlockouts
\usepackage{float}
\usepackage{amsmath,amssymb,amsfonts}
\usepackage{graphicx}
\usepackage{textcomp}
\usepackage{colortbl}
\usepackage{flushend}
\usepackage{adjustbox}
\usepackage{caption}
\usepackage{tabularx}
\usepackage{import}
\usepackage{algpseudocode,algorithm}
\usepackage{listings}
\usepackage{multirow}
\usepackage{caption}
\usepackage{subcaption}
\usepackage{hyperref}
\usepackage{circuitikz}
\usepackage{pgf}
\usepackage{pgfplots}
\usepackage{xcolor}
\usepackage{bm}
\usepackage{pgfplotstable}
\pgfplotsset{compat=1.8}
\usepackage{adjustbox}
\usepackage{changepage}
\usepackage{dblfloatfix}
\usepackage[frozencache,cachedir=.]{minted} 
\usepackage{xpatch}

\usepackage[utf8]{inputenc}

\newcommand\xqed[1]{%
  \leavevmode\unskip\penalty9999 \hbox{}\nobreak\hfill
  \quad\hbox{#1}}
\newcommand\exampleEnd{\xqed{$\blacksquare$}}

\definecolor{bblue}{HTML}{235996}
\definecolor{rred}{HTML}{99004D}
\definecolor{ggreen}{HTML}{417505}
\definecolor{ppurple}{HTML}{7859A3}

\makeatletter

\tikzstyle{chart}=[
    legend label/.style={font={\scriptsize},anchor=west,align=left},
    legend box/.style={rectangle, draw, minimum size=5pt},
    axis/.style={black,semithick,->},
    axis label/.style={anchor=east,font={\tiny}},
]

\tikzstyle{bar chart}=[
    chart,
    bar width/.code={
        \pgfmathparse{##1/2}
        \global\let\bar@w\pgfmathresult
    },
    bar/.style={very thick, draw=white},
    bar label/.style={font={\bf\small},anchor=north},
    bar value/.style={font={\footnotesize}},
    bar width=.75,
]

\tikzstyle{pie chart}=[
    chart,
    slice/.style={line cap=round, line join=round, very thick,draw=white},
    pie title/.style={font={\bf},shift={(0cm,-0.2cm)}},
    slice type/.style 2 args={
        ##1/.style={fill=##2},
        values of ##1/.style={}
    }
]

\pgfdeclarelayer{background}
\pgfdeclarelayer{foreground}
\pgfsetlayers{background,main,foreground}

\newcommand{\pie}[3][]{
    \begin{scope}[#1]
    \pgfmathsetmacro{\curA}{90}
    \pgfmathsetmacro{\r}{1}
    \def\c{(0,0)}
    \node[pie title] at (90:1.3) {#2};
    \foreach \v/\s in{#3}{
        \pgfmathsetmacro{\deltaA}{\v/100*360}
        \pgfmathsetmacro{\nextA}{\curA + \deltaA}
        \pgfmathsetmacro{\midA}{(\curA+\nextA)/2}

        \path[slice,\s] \c
            -- +(\curA:\r)
            arc (\curA:\nextA:\r)
            -- cycle;
        \pgfmathsetmacro{\d}{max((\deltaA * -(.5/50) + 1) , .5)}

        \begin{pgfonlayer}{foreground}
        \path \c -- node[pos=\d,pie values,values of \s]{$\v\%$} +(\midA:\r);
        \end{pgfonlayer}

        \global\let\curA\nextA
    }
    \end{scope}
}

 \newcounter{mycounter}

\def\thesubsubsectiondis{\unskip\arabic{subsubsection})}
\def\BibTeX{{\rm B\kern-.05em{\sc i\kern-.025em b}\kern-.08em
    T\kern-.1667em\lower.7ex\hbox{E}\kern-.125emX}}

\newcolumntype{Y}{>{\centering\arraybackslash}X}
    
\usepackage{paralist}

\begin{document}

\newenvironment{code}{\captionsetup{type=listing}}{}


\newlength{\savejot}
\setlength{\savejot}{\jot}

\newcommand{\xor}{{\,\oplus\,}}
\newcommand{\bits}{{\{0,1\}}}

\newcommand{\heading}[1]{{\vspace{10pt}\noindent{\textsc{#1}}}}
\newcommand{\noskipheading}[1]{{\noindent{\textsc{#1}}}}

\newcommand{\headingg}[1]{{\textsc{#1}}}
\newcommand{\Heading}[1]{{\vspace{8pt}\noindent\textbf{#1}}}

\newcommand{\emptystring}{\varepsilon}
\newcommand{\concat}{\:\|\:}
\newcommand{\Concat}{\;\|\;}
\newcommand{\Colon}{{\,:\;\,}}

\newcommand{\N}{{{\sf N}}}
\newcommand{\R}{{{\rm\bf R}}}
\newcommand{\Y}{{{\sf Y}}}
\newcommand{\Z}{{{Z}}}

\newcommand{\calC}{{\cal C}}
\newcommand{\calD}{{\cal D}}
\newcommand{\calE}{{\cal E}}
\newcommand{\calF}{{\cal F}}
\newcommand{\calI}{{\cal I}}
\newcommand{\calO}{{\cal O}}
\newcommand{\calQ}{{\cal Q}}
\newcommand{\calR}{{\cal R}}
\newcommand{\calG}{{\cal G}}
\newcommand{\calK}{{\cal K}}
\newcommand{\calN}{{\cal N}}
\newcommand{\calT}{{\cal T}}
\newcommand{\calH}{{\mathcal{H}}}
\newcommand{\calM}{{\mathcal{M}}}
\newcommand{\calV}{{\mathcal{V}}}
\newcommand{\calP}{{\mathcal{P}}}

\newcommand{\D}{{\calD}}
\newcommand{\E}{{\calE}}
\newcommand{\K}{{\calK}}
\newcommand{\I}{{\calI}}
\renewcommand{\R}{{\calR}}
\renewcommand{\O}{{\calO}}

\newcommand{\goesto}{{\rightarrow}}
\newcommand{\eqdef}{\stackrel{\rm def}{=}}
\newcommand{\getsr}{{\:\stackrel{{\scriptscriptstyle \hspace{0.2em}\$}} {\leftarrow}\:}}
\newcommand{\getsd}{{\:\stackrel{{\scriptscriptstyle \hspace{0.2em}D}} {\leftarrow}\:}}
\renewcommand{\choose}[2]{{{#1}\atopwithdelims(){#2}}}
\newcommand{\abs}[1]{{\displaystyle \left| {#1} \right| }}
\newcommand{\EE}[1]{{\E\left[{#1}\right]}}

\newcommand{\Damgard}{{\mbox{Damg{\aa}rd}}}

\newcommand{\strtonum}[1]{{ \mathsf{str2num}\left({#1}\right) }}
\newcommand{\numtostr}[2]{{ \mathsf{num2str}_{\scriptscriptstyle #2}
\left( {#1} \right) }}

\newcommand{\Adv}{{\mathbf{\bf Adv}}}

\newcommand{\Randword}{\mathrm{Rand}}
\newcommand{\Permword}{\mathrm{Perm}}
\newcommand{\Randd}[2]{\ensuremath{\Randword({#1},{#2})}}
\newcommand{\Rand}[1]{\ensuremath{\Randword({#1})}}
\newcommand{\Rn}{{\calR_n}}
\newcommand{\Pn}{{\calP_n}}
\newcommand{\Perm}[1]{\ensuremath{\Permword({#1})}}
\newcommand{\Permm}[2]{\ensuremath{\Permword({#1},{#2})}}

\newcommand{\GF}{{\mathrm{GF}}}

\newcommand{\then}{{;\;}}
\newcommand{\andthen}{{\::\;\;}}

\newcommand{\ie}{i.e.}
\newcommand{\eg}{e.g.}
\newcommand{\adv}{\ensuremath{\mathrm{Adv}}}
\newcommand{\hdir}{\mbox{\vspace{6mm}\~{ }}}

\newcolumntype{Y}{>{\centering\arraybackslash}X}

\title{\huge
HIVE: Scalable Hardware-Firmware Co-Verification using Scenario-based Decomposition and Automated Hint Extraction

\author{Aruna~Jayasena,~\IEEEmembership{Student Member,~IEEE,}
        and~Prabhat~Mishra,~\IEEEmembership{Fellow,~IEEE,}\vspace{-0.1in}
\IEEEcompsocitemizethanks{\IEEEcompsocthanksitem A. Jayasena and P. Mishra are with the Department of Computer \& Information Science \& Engineering, University of Florida, Gainesville, Florida.}
\thanks{This work was partially supported by the grants from NSF (CCF-1908131) and Semiconductor Research Corporation (2022-HW-3128).}
} 
}

\maketitle

\pagestyle{plain}

\begin{abstract}

Hardware-firmware co-verification is critical to design trustworthy systems. 
While formal methods can provide verification guarantees, due to the complexity of firmware and hardware, it can lead to state space explosion. There are promising avenues to reduce the state space during firmware verification through manual abstraction of hardware or manual generation of hints. Manual development of abstraction or hints requires domain expertise and can be time-consuming and error-prone, leading to incorrect proofs or inaccurate results. In this paper, we effectively combine the scalability of simulation-based validation and the completeness of formal verification. Our proposed approach is applicable to actual firmware and hardware implementations without requiring any manual intervention during formal model generation or hint extraction. To reduce the state space complexity, we utilize both static module-level analysis and dynamic execution of verification scenarios to automatically generate system-level hints. These hints guide the underlying solver to perform scalable equivalence checking using proofs. The extracted hints are validated against the implementation before using them in the proofs. Experimental evaluation on RISC-V based systems demonstrates that our proposed framework is scalable due to scenario-based decomposition and automated hint extraction. Moreover, our fully automated framework can identify complex bugs in actual firmware-hardware implementations.


\end{abstract}


\section{Introduction} \label{sec:introduction}


System-on-Chip (SoC) architectures consist of components that are implemented as both hardware and firmware. External IPs such as cryptographic accelerators, and communication modules such as UART, I\textsuperscript{2}C and SPI modules use memory-mapped input/outputs (MMIO) to communicate with the processor. These devices are typically connected with the firmware applications and the device drivers. Both firmware applications and device drivers are usually written in \textit{C} language. Once the application is ready, they are compiled with the device drivers using the toolchain corresponding to the processor in the SoC. Next, a customized linker script informs the compiler about the memory configurations used in the setup to obtain the final compiled binary. An efficient and scalable framework is needed to verify such a complex system across all these layers of hardware, drivers, and applications.

There are promising test generation-based validation techniques for functional and security verification of SoCs~\cite{ahn2014automated,ahn2013modeling, feng2020p2im,jayasena2022network}. These techniques either rely on the abstraction of the hardware-firmware interactions~\cite{feng2020p2im} or applicable on specific applications~\cite{gao2022fw}. Due to exponential input space complexity, test generation based techniques are likely to miss corner cases, and therefore cannot provide a correctness guarantee. For example, we need to simulate a 64-bit adder (adds two 64-bit inputs) with $2^{128}$ test vectors to provide the verification guarantee. Clearly, it is infeasible to simulate a real-world design with all possible input vectors. 


Formal verification methods~\cite{athalye2019notary,athalye2022verifying, morozertlv,huang2018instruction,huang2018formal,huang2019ilang} can provide strong guarantees about the correctness of a system. While formal verification is not affected by input space complexity, it can lead to state space explosion when dealing with large and complex systems~\cite{jayasena2024directed}. As a result, the scalability of formal verification is limited to small parts of the design due to the fact that symbolic expression term size grows exponentially with the increasing design size. There are promising efforts for system-level formal verification that rely on the designer's expertise to guide the proofs with manual  hints~\cite{athalye2022verifying, morozertlv} to partition the state space and reduce the growth of expression term size.

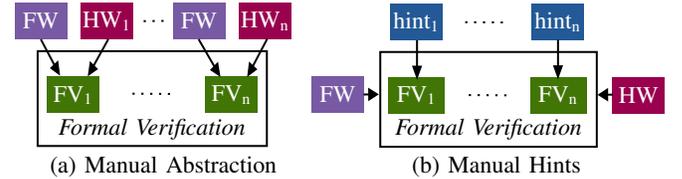
\begin{figure}[htp]
    \centering
     \vspace{-0.1in}
        \begin{subfigure}{.5\linewidth}
          \centering
          \begin{adjustwidth*}{-1.0em}{0em}
          \tikzset{every picture/.style={line width=0.75pt}} 
\small
\begin{tikzpicture}[x=0.75pt,y=0.75pt,yscale=-0.8,xscale=0.6]

\draw  [color={rgb, 255:red, 0; green, 0; blue, 0 }  ,draw opacity=1 ] (24.33,40.35) -- (217,40.35) -- (217,100.17) -- (24.33,100.17) -- cycle ;
\draw  [draw opacity=0][fill={rgb, 255:red, 153; green, 0; blue, 77 }  ,fill opacity=1 ][line width=0.75]  (60.68,9.66) -- (104.5,9.66) -- (104.5,32.87) -- (60.68,32.87) -- cycle ;

\draw  [draw opacity=0][fill={rgb, 255:red, 120; green, 89; blue, 163 }  ,fill opacity=1 ][line width=0.75]  (5.18,9.66) -- (49,9.66) -- (49,32.87) -- (5.18,32.87) -- cycle ;

\draw  [draw opacity=0][fill={rgb, 255:red, 153; green, 0; blue, 77 }  ,fill opacity=1 ][line width=0.75]  (193.68,9.66) -- (237.5,9.66) -- (237.5,32.87) -- (193.68,32.87) -- cycle ;

\draw  [draw opacity=0][fill={rgb, 255:red, 120; green, 89; blue, 163 }  ,fill opacity=1 ][line width=0.75]  (138.18,9.66) -- (182,9.66) -- (182,32.87) -- (138.18,32.87) -- cycle ;

\draw  [draw opacity=0][fill={rgb, 255:red, 65; green, 117; blue, 5 }  ,fill opacity=1 ][line width=0.75]  (32.35,56.33) -- (76.17,56.33) -- (76.17,79.54) -- (32.35,79.54) -- cycle ;

\draw  [draw opacity=0][fill={rgb, 255:red, 65; green, 117; blue, 5 }  ,fill opacity=1 ][line width=0.75]  (165.35,55.96) -- (209.17,55.96) -- (209.17,79.17) -- (165.35,79.17) -- cycle ;

\draw    (27,32.87) -- (41.47,53.42) ;
\draw [shift={(43.2,55.87)}, rotate = 234.84] [fill={rgb, 255:red, 0; green, 0; blue, 0 }  ][line width=0.08]  [draw opacity=0] (8.93,-4.29) -- (0,0) -- (8.93,4.29) -- cycle    ;
\draw    (68.18,53.85) -- (83.67,32.87) ;
\draw [shift={(66.4,56.27)}, rotate = 306.42] [fill={rgb, 255:red, 0; green, 0; blue, 0 }  ][line width=0.08]  [draw opacity=0] (8.93,-4.29) -- (0,0) -- (8.93,4.29) -- cycle    ;
\draw    (159.67,32.87) -- (176.17,54.29) ;
\draw [shift={(178,56.67)}, rotate = 232.39] [fill={rgb, 255:red, 0; green, 0; blue, 0 }  ][line width=0.08]  [draw opacity=0] (8.93,-4.29) -- (0,0) -- (8.93,4.29) -- cycle    ;
\draw    (202.12,54.21) -- (217.2,32.67) ;
\draw [shift={(200.4,56.67)}, rotate = 304.99] [fill={rgb, 255:red, 0; green, 0; blue, 0 }  ][line width=0.08]  [draw opacity=0] (8.93,-4.29) -- (0,0) -- (8.93,4.29) -- cycle    ;
\draw  [dash pattern={on 0.84pt off 2.51pt}]  (104,67.17) -- (140,67.34) ;
\draw  [dash pattern={on 0.84pt off 2.51pt}]  (113.4,21.4) -- (131.2,21.24) ;

\draw (82.59,21.24) node   [align=left] {\begin{minipage}[lt]{29.8pt}\setlength\topsep{0pt}
\begin{center}
\textcolor[rgb]{1,1,1}{HW}\textcolor[rgb]{1,1,1}{\textsubscript{1}}
\end{center}

\end{minipage}};
\draw (120.42,89.68) node  [color={rgb, 255:red, 0; green, 0; blue, 0 }  ,opacity=1 ] [align=left] {\begin{minipage}[lt]{131.35pt}\setlength\topsep{0pt}
\begin{center}
\textit{Formal Verification}
\end{center}

\end{minipage}};
\draw (27.09,21.24) node   [align=left] {\begin{minipage}[lt]{29.8pt}\setlength\topsep{0pt}
\begin{center}
\textcolor[rgb]{1,1,1}{FW}
\end{center}

\end{minipage}};
\draw (160.09,21.24) node   [align=left] {\begin{minipage}[lt]{29.8pt}\setlength\topsep{0pt}
\begin{center}
\textcolor[rgb]{1,1,1}{FW}
\end{center}

\end{minipage}};
\draw (215.59,21.24) node   [align=left] {\begin{minipage}[lt]{29.8pt}\setlength\topsep{0pt}
\begin{center}
\textcolor[rgb]{1,1,1}{HW}\textcolor[rgb]{1,1,1}{\textsubscript{n}}
\end{center}

\end{minipage}};
\draw (54.26,67.91) node   [align=left] {\begin{minipage}[lt]{29.8pt}\setlength\topsep{0pt}
\begin{center}
\textcolor[rgb]{1,1,1}{FV}\textcolor[rgb]{1,1,1}{\textsubscript{1}}
\end{center}

\end{minipage}};
\draw (187.26,67.54) node   [align=left] {\begin{minipage}[lt]{29.8pt}\setlength\topsep{0pt}
\begin{center}
\textcolor[rgb]{1,1,1}{FV}\textcolor[rgb]{1,1,1}{\textsubscript{n}}
\end{center}

\end{minipage}};

\end{tikzpicture}
          \end{adjustwidth*}
          \vspace{-0.05in}
          \caption{Manual Abstraction}
          \label{fig:avenue1}
        \end{subfigure}%
        \begin{subfigure}{.5\linewidth}
          \centering
          \begin{adjustwidth*}{-1.5em}{-4em}
          \tikzset{every picture/.style={line width=0.75pt}} 
\small
\begin{tikzpicture}[x=0.75pt,y=0.75pt,yscale=-0.8,xscale=0.6]

\draw  [draw opacity=0][fill={rgb, 255:red, 153; green, 0; blue, 77 }  ,fill opacity=1 ][line width=0.75]  (257.85,56) -- (301.67,56) -- (301.67,79.2) -- (257.85,79.2) -- cycle ;

\draw  [draw opacity=0][fill={rgb, 255:red, 120; green, 89; blue, 163 }  ,fill opacity=1 ][line width=0.75]  (4.85,55.66) -- (48.67,55.66) -- (48.67,78.87) -- (4.85,78.87) -- cycle ;

\draw  [color={rgb, 255:red, 0; green, 0; blue, 0 }  ,draw opacity=1 ] (62.6,40.35) -- (244.6,40.35) -- (244.6,100.17) -- (62.6,100.17) -- cycle ;
\draw  [draw opacity=0][fill={rgb, 255:red, 65; green, 117; blue, 5 }  ,fill opacity=1 ][line width=0.75]  (188.6,55.96) -- (235.83,55.96) -- (235.83,79.17) -- (188.6,79.17) -- cycle ;

\draw  [draw opacity=0][fill={rgb, 255:red, 35; green, 89; blue, 150 }  ,fill opacity=1 ][line width=0.75]  (71.35,9.66) -- (115.17,9.66) -- (115.17,32.87) -- (71.35,32.87) -- cycle ;

\draw  [draw opacity=0][fill={rgb, 255:red, 35; green, 89; blue, 150 }  ,fill opacity=1 ][line width=0.75]  (190.01,9.66) -- (233.83,9.66) -- (233.83,32.87) -- (190.01,32.87) -- cycle ;

\draw    (58.4,67.09) -- (48.6,67.17) ;
\draw [shift={(61.4,67.07)}, rotate = 179.56] [fill={rgb, 255:red, 0; green, 0; blue, 0 }  ][line width=0.08]  [draw opacity=0] (8.93,-4.29) -- (0,0) -- (8.93,4.29) -- cycle    ;
\draw    (258,67.27) -- (248.2,67.34) ;
\draw [shift={(245.2,67.37)}, rotate = 359.56] [fill={rgb, 255:red, 0; green, 0; blue, 0 }  ][line width=0.08]  [draw opacity=0] (8.93,-4.29) -- (0,0) -- (8.93,4.29) -- cycle    ;
\draw    (212,32.67) -- (212,54.67) ;
\draw [shift={(212,57.67)}, rotate = 270] [fill={rgb, 255:red, 0; green, 0; blue, 0 }  ][line width=0.08]  [draw opacity=0] (8.93,-4.29) -- (0,0) -- (8.93,4.29) -- cycle    ;
\draw  [dash pattern={on 0.84pt off 2.51pt}]  (137,67.54) -- (173,67.71) ;
\draw  [dash pattern={on 0.84pt off 2.51pt}]  (134,21.54) -- (170,21.71) ;
\draw  [draw opacity=0][fill={rgb, 255:red, 65; green, 117; blue, 5 }  ,fill opacity=1 ][line width=0.75]  (69.68,55.96) -- (116.92,55.96) -- (116.92,79.17) -- (69.68,79.17) -- cycle ;

\draw    (93.5,32.92) -- (93.5,54.92) ;
\draw [shift={(93.5,57.92)}, rotate = 270] [fill={rgb, 255:red, 0; green, 0; blue, 0 }  ][line width=0.08]  [draw opacity=0] (8.93,-4.29) -- (0,0) -- (8.93,4.29) -- cycle    ;

\draw (26.76,67.24) node   [align=left] {\begin{minipage}[lt]{29.8pt}\setlength\topsep{0pt}
\begin{center}
\textcolor[rgb]{1,1,1}{FW}
\end{center}

\end{minipage}};
\draw (279.76,67.58) node   [align=left] {\begin{minipage}[lt]{29.8pt}\setlength\topsep{0pt}
\begin{center}
\textcolor[rgb]{1,1,1}{HW}
\end{center}

\end{minipage}};
\draw (212.22,67.54) node   [align=left] {\begin{minipage}[lt]{32.12pt}\setlength\topsep{0pt}
\begin{center}
\textcolor[rgb]{1,1,1}{FV}\textcolor[rgb]{1,1,1}{\textsubscript{n}}
\end{center}

\end{minipage}};
\draw (93.26,21.24) node   [align=left] {\begin{minipage}[lt]{29.8pt}\setlength\topsep{0pt}
\begin{center}
\textcolor[rgb]{1,1,1}{hint}\textcolor[rgb]{1,1,1}{\textsubscript{1}}
\end{center}

\end{minipage}};
\draw (211.92,21.24) node   [align=left] {\begin{minipage}[lt]{29.8pt}\setlength\topsep{0pt}
\begin{center}
\textcolor[rgb]{1,1,1}{hint}\textcolor[rgb]{1,1,1}{\textsubscript{n}}
\end{center}

\end{minipage}};
\draw (153.6,89.17) node  [color={rgb, 255:red, 0; green, 0; blue, 0 }  ,opacity=1 ] [align=left] {\begin{minipage}[lt]{124.3pt}\setlength\topsep{0pt}
\begin{center}
\textit{Formal Verification}
\end{center}

\end{minipage}};
\draw (93.3,67.54) node   [align=left] {\begin{minipage}[lt]{32.12pt}\setlength\topsep{0pt}
\begin{center}
\textcolor[rgb]{1,1,1}{FV}\textcolor[rgb]{1,1,1}{\textsubscript{1}}
\end{center}

\end{minipage}};

\end{tikzpicture}
          \end{adjustwidth*}
          \vspace{-0.05in}
          \caption{Manual Hints}
          \label{fig:avenue2}
        \end{subfigure}
      \vspace{-0.2in}
      \caption{Existing formal firmware verification avenues}
        \vspace{-0.15in}
      \label{fig:firmwareAvenues}
\end{figure}

\begin{figure*}[htp]
    \begin{center}
    \vspace{-0.2in}
        \input{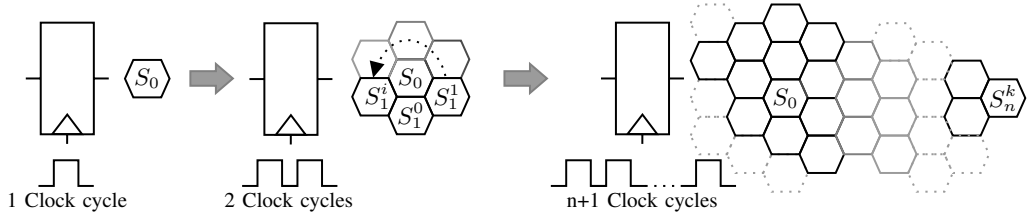}
    \end{center}
      \vspace{-0.15in}
      \caption{ State space expansion of a state register of a hardware implementation with respect to the evaluated clock cycles.}
        \vspace{-0.1in}
      \label{fig:state_expansion}
\end{figure*}

Existing firmware verification efforts explore two promising avenues to reduce the state space as illustrated in Figure~\ref{fig:firmwareAvenues}. The methods in the first category perform firmware verification on top of multiple manually abstracted versions of the hardware instead of register-transfer level (RTL) implementation. For example, Figure~\ref{fig:avenue1} shows that to perform formal verification (FV) of a  firmware (FW) for $n$ scenarios, a designer needs to manually construct the abstracted hardware model ($HW_i$) customized for the $i$-th scenario. Manual abstraction based on verification scenarios needs domain expertise and can be erroneous. Moreover, discrepancies between the actual implementation and the abstract model can lead to false positive results. The methods in the second category perform automated formal model generation for hardware, but it relies on manual extraction of hints from both firmware and hardware to guide formal proofs to avoid state explosion due to rapidly increasing term size. For example, Figure~\ref{fig:avenue2} shows that a designer needs to manually construct the hint ($hint_i$) customized for the $i$-th  scenario. Manual development of hints can be time-consuming and error-prone, leading to incorrect proofs or inaccurate results.

\vspace{-0.05in}
\subsection{State-of-the-Art in State Space Reduction}
An inherent weakness of formal verification is state space explosion, where the backend solver has too many possibilities to explore at once. 
In a system that combines both hardware and firmware, there are multiple ways that state explosion can occur. Due to the concurrency of hardware, state registers can increase rapidly in each clock cycle in which it gets evaluated. Figure~\ref{fig:state_expansion} illustrates an example scenario for a state register. When the number of unrolling cycles for the design increases, the number of possible paths that the state register can represent increases exponentially. Similarly, the number of symbolic variables also increase linearly when the implementation size increases. Moreover, firmware consists of branch statements which increases the number of possible states the system can have. These factors contribute to the symbolic state explosion and path explosion~\cite{dobrescu2015software, stoenescu2016symnet} during formal verification.

In order to guide the formal proofs without getting into the state explosion problem, there are promising state space reduction techniques. These techniques aim to reduce the size of the state space by identifying and eliminating states that are either irrelevant or equivalent to other states for a given verification context. The goal is to make the verification process more efficient and effective by reducing the number of states that need to be analyzed. We outline four popular approaches for state space reduction.

\subsubsection{Abstraction} Abstraction involves simplifying the system model by removing details that are not relevant to the properties being verified~\cite{clarke1999model}. This can include abstracting away certain variables, simplifying the system topology, or reducing the level of details in the system model. Usually, abstraction is a manual process that relies on the expertise of the verification engineer. Quality and the similarity of the abstraction determine the guarantees and accuracy of the proofs. In other words, if the abstraction does not represent the system correctly, verification can lead to false positive results.

\subsubsection{Symmetry Reduction}
Symmetry reduction involves identifying and eliminating symmetric states in the system model~\cite{clarke1999model}. Symmetric states are states that are equivalent to other states under certain transformations, such as swapping the values of variables or reordering events.

\subsubsection{Partial-order Reduction}
Partial order reduction involves analyzing only a subset of the possible orderings of events in the system model~\cite{clarke1999model}. This can be achieved by identifying independent events that can be executed in any order, and then only analyzing a relevant subset of the possible orderings.

\subsubsection{Decomposition}
Decomposition enables state-space reduction by breaking a problem into multiple sub-problems~\cite{clarke1999model}.
Instead of going through the manual and tedious task of abstraction for system validation purposes, our proposed technique divides the main verification problem into sub-problems. Then symmetric reduction and partial order reduction techniques are directly applied to each sub-problem. Our sub-problem decomposition technique adheres the basic principles of assume-guarantee~\cite{abadi1995conjoining,giannakopoulou2004assume} and compositionality~\cite{abadi1993composing,giannakopoulou2004assume}. Assume-guarantee defines the requirements that should exist in a system to decompose them into multiple sub-problems while providing the system-level guarantee. Compositionality refers to the property of a system or proof that allows it to be divided into smaller, independent parts or sub-problems, which can be analyzed or verified separately, and then combined to form a complete solution.

\begin{figure}[htp]
    \begin{center}
    \vspace{-0.1in}
    \begin{adjustwidth*}{1em}{-2.3em}
        \tikzset{every picture/.style={line width=0.75pt}} 
\small
\begin{tikzpicture}[x=0.75pt,y=0.75pt,yscale=-0.7,xscale=0.7]

\draw  [draw opacity=0][fill={rgb, 255:red, 120; green, 89; blue, 163 }  ,fill opacity=1 ][line width=0.75]  (7.19,118) -- (78.48,118) -- (78.48,145.17) -- (7.19,145.17) -- cycle ;

\draw   (-1.62,1.1) -- (203,1.1) -- (203,60.92) -- (-1.62,60.92) -- cycle ;
\draw  [draw opacity=0][fill={rgb, 255:red, 153; green, 0; blue, 77 }  ,fill opacity=1 ] (311.6,132.85) .. controls (311.6,117.77) and (340.05,105.54) .. (375.15,105.54) .. controls (410.24,105.54) and (438.69,117.77) .. (438.69,132.85) .. controls (438.69,147.93) and (410.24,160.15) .. (375.15,160.15) .. controls (340.05,160.15) and (311.6,147.93) .. (311.6,132.85) -- cycle ;

\draw [color={rgb, 255:red, 65; green, 117; blue, 5 }  ,draw opacity=1 ][line width=1.5]    (291.38,100.17) -- (86,100.17) ;
\draw [shift={(82,100.17)}, rotate = 360] [fill={rgb, 255:red, 65; green, 117; blue, 5 }  ,fill opacity=1 ][line width=0.08]  [draw opacity=0] (9.29,-4.46) -- (0,0) -- (9.29,4.46) -- cycle    ;
\draw  [draw opacity=0][fill={rgb, 255:red, 120; green, 89; blue, 163 }  ,fill opacity=1 ][line width=0.75]  (7.18,28.66) -- (78.39,28.66) -- (78.39,55.84) -- (7.18,55.84) -- cycle ;

\draw  [color={rgb, 255:red, 65; green, 117; blue, 5 }  ,draw opacity=1 ][line width=1.5]  (291.8,68.67) -- (459,68.67) -- (459,169.63) -- (291.8,169.63) -- cycle ;
\draw   (210.6,1.17) -- (459,1.17) -- (459,60.92) -- (210.6,60.92) -- cycle ;
\draw   (-0.82,109.5) -- (283.8,109.5) -- (283.8,169.63) -- (-0.82,169.63) -- cycle ;
\draw  [color={rgb, 255:red, 208; green, 2; blue, 27 }  ,draw opacity=1 ][line width=1.5]  (3,22.17) -- (82.5,22.17) -- (82.5,151.5) -- (3,151.5) -- cycle ;
\draw  [draw opacity=0][fill={rgb, 255:red, 120; green, 89; blue, 163 }  ,fill opacity=1 ][line width=0.75]  (218.65,29) -- (290.57,29) -- (290.57,56.17) -- (218.65,56.17) -- cycle ;

\draw  [draw opacity=0][fill={rgb, 255:red, 120; green, 89; blue, 163 }  ,fill opacity=1 ][line width=0.75]  (380.08,28.33) -- (452,28.33) -- (452,55.51) -- (380.08,55.51) -- cycle ;

\draw [line width=0.75]  [dash pattern={on 4.5pt off 4.5pt}]  (184.07,60.92) .. controls (184.12,97.12) and (238.98,89.03) .. (288.41,88.78) ;
\draw [shift={(290.67,88.78)}, rotate = 180] [fill={rgb, 255:red, 0; green, 0; blue, 0 }  ][line width=0.08]  [draw opacity=0] (8.93,-4.29) -- (0,0) -- (8.93,4.29) -- cycle    ;
\draw  [draw opacity=0][fill={rgb, 255:red, 120; green, 89; blue, 163 }  ,fill opacity=1 ][line width=0.75]  (124.67,28.33) -- (196.06,28.33) -- (196.06,55.51) -- (124.67,55.51) -- cycle ;

\draw  [draw opacity=0][fill={rgb, 255:red, 120; green, 89; blue, 163 }  ,fill opacity=1 ][line width=0.75]  (89.88,118.33) -- (161.17,118.33) -- (161.17,145.51) -- (89.88,145.51) -- cycle ;

\draw [line width=0.75]  [dash pattern={on 4.5pt off 4.5pt}]  (232.8,60.92) .. controls (232.87,78.86) and (257.57,77.99) .. (287.69,77.24) ;
\draw [shift={(290.5,77.17)}, rotate = 178.63] [fill={rgb, 255:red, 0; green, 0; blue, 0 }  ][line width=0.08]  [draw opacity=0] (8.93,-4.29) -- (0,0) -- (8.93,4.29) -- cycle    ;
\draw  [draw opacity=0][fill={rgb, 255:red, 120; green, 89; blue, 163 }  ,fill opacity=1 ][line width=0.75]  (204.18,117.98) -- (275.47,117.98) -- (275.47,145.15) -- (204.18,145.15) -- cycle ;

\draw (375.2,85.95) node   [align=left] {\begin{minipage}[lt]{113.97pt}\setlength\topsep{0pt}
\begin{center}
\textbf{\textcolor[rgb]{0.25,0.46,0.02}{Supporting Artifacts}}
\end{center}

\end{minipage}};
\draw (375.6,136.85) node   [align=left] {\begin{minipage}[lt]{85.81pt}\setlength\topsep{0pt}
\begin{center}
\textcolor[rgb]{1,1,1}{System Level}\\\textcolor[rgb]{1,1,1}{Analysis}
\end{center}

\end{minipage}};
\draw (42.79,42.22) node   [align=left] {\begin{minipage}[lt]{48.43pt}\setlength\topsep{0pt}
\begin{center}
\textcolor[rgb]{1,1,1}{main.c}
\end{center}

\end{minipage}};
\draw (101.09,10.58) node   [align=left] {\begin{minipage}[lt]{138.6pt}\setlength\topsep{0pt}
\begin{center}
\textit{Firmware and Drivers}
\end{center}

\end{minipage}};
\draw (334.8,10.42) node   [align=left] {\begin{minipage}[lt]{168.91pt}\setlength\topsep{0pt}
\begin{center}
\textit{Hardware}
\end{center}

\end{minipage}};
\draw (141.29,159.13) node   [align=left] {\begin{minipage}[lt]{193.27pt}\setlength\topsep{0pt}
\begin{center}
\textit{Specifications}
\end{center}

\end{minipage}};
\draw (42.83,131.56) node   [align=left] {\begin{minipage}[lt]{48.48pt}\setlength\topsep{0pt}
\begin{center}
\textcolor[rgb]{1,1,1}{main.{\footnotesize spec}}
\end{center}

\end{minipage}};
\draw (254.61,42.56) node   [align=left] {\begin{minipage}[lt]{48.91pt}\setlength\topsep{0pt}
\begin{center}
\textcolor[rgb]{1,1,1}{cpu.v}
\end{center}

\end{minipage}};
\draw (416.04,41.89) node   [align=left] {\begin{minipage}[lt]{48.91pt}\setlength\topsep{0pt}
\begin{center}
\textcolor[rgb]{1,1,1}{rom.v}
\end{center}

\end{minipage}};
\draw (109.85,44.07) node   [align=left] {\begin{minipage}[lt]{25.36pt}\setlength\topsep{0pt}
.... ..
\end{minipage}};
\draw (335,43.27) node   [align=left] {\begin{minipage}[lt]{26.93pt}\setlength\topsep{0pt}
.... ..
\end{minipage}};
\draw (188.37,132.17) node   [align=left] {\begin{minipage}[lt]{25.33pt}\setlength\topsep{0pt}
.... ..
\end{minipage}};
\draw (160.37,41.89) node   [align=left] {\begin{minipage}[lt]{48.55pt}\setlength\topsep{0pt}
\begin{center}
\textcolor[rgb]{1,1,1}{driver.c}
\end{center}

\end{minipage}};
\draw (125.52,131.89) node   [align=left] {\begin{minipage}[lt]{48.48pt}\setlength\topsep{0pt}
\begin{center}
\textcolor[rgb]{1,1,1}{driver.{\footnotesize spec}}
\end{center}

\end{minipage}};
\draw (167.12,90.32) node  [rotate=-359.67] [align=left] {\begin{minipage}[lt]{93.57pt}\setlength\topsep{0pt}
\begin{center}
\textbf{\textcolor[rgb]{0.25,0.46,0.02}{Hints}}
\end{center}

\end{minipage}};
\draw (239.82,131.53) node   [align=left] {\begin{minipage}[lt]{48.48pt}\setlength\topsep{0pt}
\begin{center}
\textcolor[rgb]{1,1,1}{cpu.{\footnotesize spec}}
\end{center}

\end{minipage}};
\draw (43.1,87.12) node   [align=left] {\begin{minipage}[lt]{51.54pt}\setlength\topsep{0pt}
\begin{center}
\textcolor[rgb]{0.82,0.01,0.11}{\textbf{Proof}}
\end{center}

\end{minipage}};

\end{tikzpicture}
    \end{adjustwidth*}
    \end{center}
      \vspace{-0.15in}
      \caption{Overview of our automated \textit{\underline{Hi}nt-based \underline{Ve}rification} (\textit{HIVE}) framework. Formal proofs are supported by hints.}
        \vspace{-0.2in}
      \label{fig:Intro}
\end{figure}
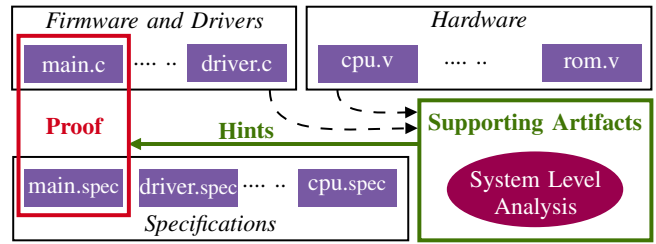

\subsection{Research Contributions}

In this paper, we effectively combine the rigor and completeness of formal verification and the scalability of test generation based validation.  We simplify the system validation problem by dividing it into sub-problems. For each sub-problem, we check the consistency of the implementation against the specification. This covers input output consistency of sub-problems and ensures that the implementation satisfies the specification. To further reduce the state space for each sub-problem, we utilize both concrete simulation and static analysis of the designs to extract proof supporting artifacts. Figure~\ref{fig:Intro} presents an overview of our proposed framework, \textit{HIVE} (\textit{\underline{Hi}nt-based \underline{Ve}rification}), where proof supporting artifacts required for sub-problems are extracted through static analysis of the implementation as well as simulation of relevant test scenarios. 

\begin{figure}[htp]
     \vspace{-0.10in}
  
    \begin{subfigure}{1\linewidth}
      \centering
        \tikzset{every picture/.style={line width=0.75pt}} 

\begin{tikzpicture}[x=0.75pt,y=0.75pt,yscale=-0.6,xscale=0.95]

\draw  [draw opacity=0][fill={rgb, 255:red, 153; green, 0; blue, 77 }  ,fill opacity=1 ] (286.6,250.94) .. controls (286.6,240.5) and (298.51,232.03) .. (313.2,232.03) .. controls (327.89,232.03) and (339.8,240.5) .. (339.8,250.94) .. controls (339.8,261.38) and (327.89,269.84) .. (313.2,269.84) .. controls (298.51,269.84) and (286.6,261.38) .. (286.6,250.94) -- cycle ;

\draw   (280.57,197.53) -- (347.43,197.53) -- (347.43,317.1) -- (280.57,317.1) -- cycle ;
\draw   (72.8,198.1) -- (142.08,198.1) -- (142.08,317.67) -- (72.8,317.67) -- cycle ;
\draw  [draw opacity=0][fill={rgb, 255:red, 120; green, 89; blue, 163 }  ,fill opacity=1 ][line width=0.75]  (90.25,202.99) -- (124.29,202.99) -- (124.29,230.42) -- (90.25,230.42) -- cycle ;

\draw  [draw opacity=0][fill={rgb, 255:red, 120; green, 89; blue, 163 }  ,fill opacity=1 ][line width=0.75]  (90.24,280.99) -- (124.19,280.99) -- (124.19,309.42) -- (90.24,309.42) -- cycle ;

\draw    (124.29,216.26) -- (212.71,215.96) ;
\draw [shift={(215.71,215.95)}, rotate = 179.81] [fill={rgb, 255:red, 0; green, 0; blue, 0 }  ][line width=0.08]  [draw opacity=0] (8.93,-4.29) -- (0,0) -- (8.93,4.29) -- cycle    ;
\draw    (124.29,295.24) -- (212.43,294.69) ;
\draw [shift={(215.43,294.67)}, rotate = 179.64] [fill={rgb, 255:red, 0; green, 0; blue, 0 }  ][line width=0.08]  [draw opacity=0] (8.93,-4.29) -- (0,0) -- (8.93,4.29) -- cycle    ;
\draw    (248.86,208.24) -- (313.2,207.9) -- (313.38,229.18) ;
\draw [shift={(313.4,232.18)}, rotate = 269.53] [fill={rgb, 255:red, 0; green, 0; blue, 0 }  ][line width=0.08]  [draw opacity=0] (8.93,-4.29) -- (0,0) -- (8.93,4.29) -- cycle    ;
\draw    (249.4,291.4) -- (313.4,291.4) -- (313.21,272.84) ;
\draw [shift={(313.18,269.84)}, rotate = 89.42] [fill={rgb, 255:red, 0; green, 0; blue, 0 }  ][line width=0.08]  [draw opacity=0] (8.93,-4.29) -- (0,0) -- (8.93,4.29) -- cycle    ;
\draw  [draw opacity=0][fill={rgb, 255:red, 120; green, 89; blue, 163 }  ,fill opacity=1 ][line width=0.75]  (215.39,202.42) -- (249.43,202.42) -- (249.43,229.85) -- (215.39,229.85) -- cycle ;

\draw  [draw opacity=0][fill={rgb, 255:red, 120; green, 89; blue, 163 }  ,fill opacity=1 ][line width=0.75]  (215.39,280.42) -- (249.34,280.42) -- (249.34,308.85) -- (215.39,308.85) -- cycle ;

\draw   (197.94,197.53) -- (267.23,197.53) -- (267.23,317.1) -- (197.94,317.1) -- cycle ;

\draw (107.35,257.88) node   [align=left] {\begin{minipage}[lt]{46.51pt}\setlength\topsep{0pt}
\begin{center}
\textit{Original Models}
\end{center}

\end{minipage}};
\draw (232.17,257.88) node   [align=left] {\begin{minipage}[lt]{47.02pt}\setlength\topsep{0pt}
\begin{center}
\textit{Formal}\\\textit{Models}
\end{center}

\end{minipage}};
\draw (312.84,300) node   [align=left] {\begin{minipage}[lt]{45.49pt}\setlength\topsep{0pt}
\begin{center}
\textit{Verification}
\end{center}

\end{minipage}};
\draw (169.75,207.6) node   [align=left] {\begin{minipage}[lt]{36.38pt}\setlength\topsep{0pt}
\begin{center}
{\small Manual}
\end{center}

\end{minipage}};
\draw (169.94,285.88) node   [align=left] {\begin{minipage}[lt]{36.41pt}\setlength\topsep{0pt}
\begin{center}
{\small Manual}
\end{center}

\end{minipage}};
\draw (232.36,294.6) node   [align=left] {\begin{minipage}[lt]{23.09pt}\setlength\topsep{0pt}
\begin{center}
\textcolor[rgb]{1,1,1}{{\large S'}}
\end{center}

\end{minipage}};
\draw (232.41,216.1) node   [align=left] {\begin{minipage}[lt]{23.14pt}\setlength\topsep{0pt}
\begin{center}
\textcolor[rgb]{1,1,1}{{\large I'}}
\end{center}

\end{minipage}};
\draw (107.22,295.18) node   [align=left] {\begin{minipage}[lt]{23.09pt}\setlength\topsep{0pt}
\begin{center}
\textcolor[rgb]{1,1,1}{{\large S}}
\end{center}

\end{minipage}};
\draw (107.27,216.68) node   [align=left] {\begin{minipage}[lt]{23.14pt}\setlength\topsep{0pt}
\begin{center}
\textcolor[rgb]{1,1,1}{{\large I}}
\end{center}

\end{minipage}};
\draw (313.39,250.94) node   [align=left] {\begin{minipage}[lt]{35.92pt}\setlength\topsep{0pt}
\begin{center}
\textcolor[rgb]{1,1,1}{I' $\subseteq$ S'}
\end{center}

\end{minipage}};

\end{tikzpicture}
      \vspace{-0.05in}
      \caption{Existing approaches suffer from state explosion}
      \label{fig:existing}
    \end{subfigure}%

    \begin{subfigure}{1\linewidth}
      \centering
      
      \input{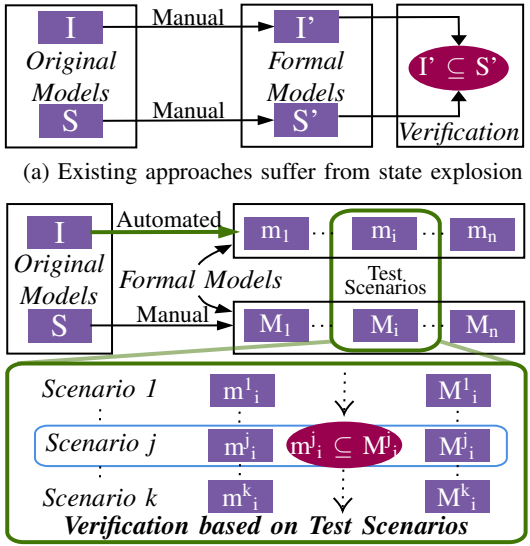}
      \vspace{-0.05in}
      \caption{Proposed method utilizes automated scenario-based analysis. }
      \label{fig:proposedscenariobased}

    \end{subfigure}%

      \caption{\textit{HIVE} utilizes both module-level and test scenario-based decomposition for improving the scalability of proofs.}
        \vspace{-0.1in}
      \label{fig:contribution}
\end{figure}

Figure~\ref{fig:contribution} highlights the novelty of our methodology compared to the existing approaches. Figure~\ref{fig:existing} shows that the existing methods rely on manual abstraction of the hardware implementation, but it can lead to false positive results. As shown in Figure~\ref{fig:proposedscenariobased}, our proposed approach has two objectives: (i) automated formal model generation to avoid false positive results, and (ii) scenario-based verification to reduce the state space specific to the verification scenarios. Here, $M_i$ and $m_i$ represent the $i$-th component (module) in the specification (S) and implementation (I), respectively. Similarly, $M^j_i$ and $m^j_i$ represent the $i$-th module customized for the $j$-th scenario in the specification and implementation, respectively.

\noindent Specifically, this paper makes the following contributions. 

\begin{itemize}
    \item It enables automated generation of formal models from the system-level implementation. We compile the firmware and combine it with the hardware to obtain the system model. 

    \item To avoid state space explosion, it decomposes the system-level equivalence checking problem into sub-module level equivalence checking problems. We utilize function-level boundaries of firmware and module-level boundaries of hardware to perform module-level equivalence checking.

    \item We utilize both architectural and behavioral models through static analysis of modules and dynamic analysis of the implementation. This leads to the automated generation of module interactions and supporting artifacts.

    \item It reduces the state space further by scenario-based analysis. We utilize supporting artifacts to generate system-level hints and perform path prioritization and symbolic state reduction related to the scenario under validation. The generated hints guide the equivalence-checking proofs in the simplified state space specific to the scenario as illustrated in Figure~\ref{fig:proposedscenariobased}.

\end{itemize}
 
{Our proposed framework leverages the strengths of both formal verification and simulation-based validation for scalable validation of firmware running on hardware (RTL) implementation.  The design slicing feature of concrete simulation-based validation is combined with symbolic simulation to formally evaluate the implementation against target verification scenarios.} Since \textit{HIVE} generates the formal model automatically from the implementation, it is a complete model by construction relative to the implementation. Any issue found during  verification can be corrected directly on the implementation and re-validated using our proposed method. This is in contrast with the existing methods that require fixes and re-validation of multiple (abstract and implementation) models.


The rest of the paper is organized as follows, First, we discuss the background and related work. Next, we discuss the \textit{HIVE} methodology in detail. Then, we present experimental results with three case studies. Finally, we discuss the applicability and limitations of the proposed framework.

\begin{figure*}[htp]
    \begin{center}
    \vspace{-0.2in}
        \small
        \input{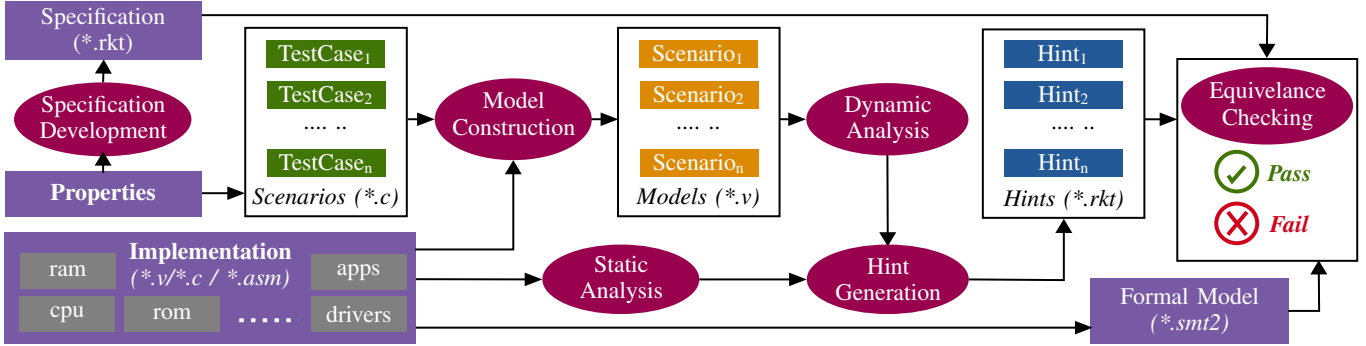}
    \end{center}
      \vspace{-0.15in}
      \caption{Overview of \textit{HIVE} framework that consists of five main steps:  outlining test scenarios, system model construction, supporting artifact generation by static and dynamic analysis, hint generation, and equivalence checking.}
    \vspace{-0.1in}
      \label{fig:overview}
\end{figure*}

\section{Background and Related Work}
\label{relatedwork}

In this section, we first survey formal methods that utilize manual abstraction or manual generation of hints for firmware verification. Next, we discuss firmware validation methods using concolic testing.

\vspace{-0.1in}
\subsection{Verification using Manual Abstraction}

Blom et al. have identified overlapping states in a formal model based on the cone of confluence~\cite{blom2002state}. The authors propose a prioritization algorithm to identify and detect confluence properties of the models under verification. Yorav et al. have utilized the control-flow graph to simplify the transition diagram for validation of parallel programs~\cite{yorav2004static}. 
Vasudevan et al. have reduced the state space through antecedent-conditioned slicing for hardware designs~\cite{vasudevan2005efficient}. Their technique reduces the state space through abstraction, and it can be utilized for the verification of safety properties in the form of $G(A \Rightarrow C)$, where $A$ (antecedent) and $C$ (consequent) are propositional logic formulas. An improved version of the algorithm utilizes the extra information from the antecedent to prune and limit the state space further for a particular problem~\cite{vasudevan2007improved}.
Analysis of Verilog designs with transaction-like protocol descriptions to reduce the state space for verification of protocols is discussed in~\cite{stuber2017formal}. The authors consider different SystemVerilog constructs with protocol-targeted assumptions for the abstraction of design and define and limit the state space of the verification problem. Another framework to validate \textit{SystemC} models utilizing a \textit{SystemC} intermediate representation is proposed in~\cite{herber2014rescue}. The authors perform static analysis on formal models and perform model slicing with abstraction to reduce the state space. None of these approaches consider state space reduction in the context of hardware-firmware verification.

Instruction Level Abstraction (ILA) is a popular technique used to formally verify firmware in System-on-Chip (SoC) designs~\cite{huang2018instruction,huang2018formal,huang2019ilang,subramanyan2015template, zhang2018ila}. ILA abstracts a processor design at the instruction level for verification purposes, which allows the verification of the design to be performed at a higher level of abstraction than the gate-level or RTL-level implementation. ILA involves modeling the processor's behavior as a set of instructions that can be executed by the processor, rather than modeling the behavior of the individual gates and wires that make up the processor. ILA can also enable the effective reuse of verification efforts across different processor designs, as the instruction-level model can be reused for different implementations of the same processor architecture. Note that ILA generation is a manual and tedious task that requires a thorough understanding of the particular processor design. Moreover,  incorrect ILA will lead to inaccurate results. Furthermore, in order to verify external accelerators and modules that connects with the SoC, separate abstraction for their implementations are required (similar to Figure~\ref{fig:avenue1}).

\subsection{Formal Verification using Manual Hints}
\textit{RTLV}~\cite{morozertlv} is a formal verification framework that utilizes hybrid symbolic execution to verify firmware that runs on the hardware. It utilizes \textit{Rosette} solver-aided programming language to perform the symbolic evaluation. Compared to the existing frameworks, it can handle complex RISC-V-based CPUs with unrolling over 20,000 clock cycles. In order to guide the proofs, the authors have used manually developed hints that are specific to the design under verification (similar to Figure~\ref{fig:avenue2}).
Athalye et al. have developed a framework for validation of hardware security modules (HSM) with information refinement~\cite{athalye2022verifying}. The authors have performed equivalence checking on three simple HSM designs implemented with both hardware and firmware. The proposed technique has several limitations. To avoid the state explosion, the authors have manually provided hints to reduce the rapidly growing symbolic term space. Due to the fact that the verification problem is posed as one verification problem to the underlying solver, complex design aspects such as cryptographic calculations cannot be handled with the proposed technique due to solver timeouts.

The existing approaches rely on manual hints~\cite{athalye2022verifying,morozertlv}, which requires significant expertise and can be error-prone. An incorrect hint would add more complexity to the solver in terms of state space and will eventually fail. This highlights the need for automated generation of proof supporting artifacts as well as system-level hints for scalable firmware validation.

\subsection{Firmware Validation using Concolic Testing}

Ahn et al. presented a concolic testing framework based on the concept of consumer-producer relationships between hardware and firmware~\cite{ahn2014automated}. Their technique combines the transaction-level models (TLM) of the implementation, which are implemented in \textit{SystemC}, with the firmware implemented in \textit{C} for the evaluation. The symbolic execution is performed with KLEE~\cite{cadar2008klee}. The authors remove the concurrent nature of the firmware by modeling the interactions as sequential events through mapping interactions to the consumer and producer models. Concolic testing approach based on virtual prototypes for verification of hardware and software models is proposed in~\cite{chen2019end}. The authors utilize virtual device models as the hardware model and run the software on the virtual device model while collecting traces. These traces are then utilized to guide the test generation process in the required direction. Although the concolic testing framework is effective in generating test cases to activate corner cases, these techniques also require virtual prototype models of the hardware implementations. Note that the generated tests are abstract tests, which may not be directly applicable to actual implementation. Moreover, the virtual prototype model needs to be in a form that can be analyzed with existing concolic testing tools that were developed for software verification. Therefore, these approaches are not suitable for firmware validation on top of hardware (RTL) implementation.


There are promising efforts that enable concolic testing on RTL-level hardware implementations~\cite{lyu2020scalable,ahmed2018directed,lyu2019automated}. These techniques operate on the RTL-level hardware models and the applicability is based on the controllability of the design with respect to the inputs of the designs. A major limitation of applying these techniques for hardware-firmware validation is that most of the hardware-firmware interactions are controlled through the firmware. Therefore, concolic testing needs to generate and modify the parts of the firmware, which is infeasible due to the complexity and the device-specific nature (memory layout and instruction set) of the firmware. 



\section{HIVE: Automated and Scalable Hint-based Formal Verification}\label{sec:methodology}

\begin{figure}[htp]
    \begin{center}
        \footnotesize
        \input{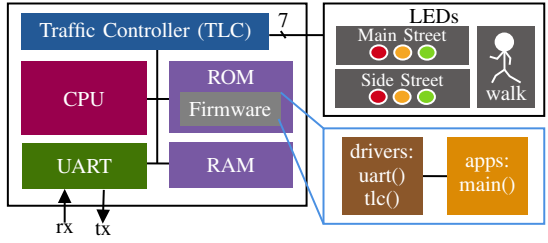}
    \end{center}
      \vspace{-0.1in}
      \caption{SoC used for a traffic light controller. }
    \vspace{-0.1in}
      \label{fig:TLC}
\end{figure}

Figure~\ref{fig:overview} provides an overview of our proposed automated \underline{Hi}nt-based \underline{Ve}rification (\textit{HIVE}) framework.
The major steps of the framework are outlined in Algorithm~\ref{algo:HIVE}.
First, we need to consider the verification scenarios and associated tests to cover these scenarios, this is the input to the algorithm. Next, the firmware tests are compiled and corresponding binaries are obtained (Line 1-7). Then, we construct the system model (Line 8).
Next, we extract the proof supporting artifacts from the designs using dynamic analysis and perform automated hint extraction (Line 9-15). Finally, for each simplified sub-problem, we utilize the generated hints to perform formal verification (Line 16-21). The remainder of this section describes each step in detail. We use the SoC implementation shown in Figure~\ref{fig:TLC} to describe each step of Algorithm~\ref{algo:HIVE} using illustrative examples.

\vspace{0.05in}
\noindent \underline{{Example 1 (SoC)}}:
\textit{Consider a firmware configurable traffic light controller design implemented on a RISC-V-instruction-set-based SoC, as shown in Figure~\ref{fig:TLC}. The SoC design is implemented in Verilog and consists of CPU, ROM, RAM, UART, and TLC modules as illustrated in the figure. In this example, firmware is able to control timing parameter values for different states of the traffic light controller, and corresponding state machine transitions are defined in the hardware description. The traffic light controller is connected to the processor of the SoC with memory-mapped I/O. There are two separate drivers for controlling the UART and TLC from the firmware. 
Further, all the sensor data is communicated to the SoC via an external device via the UART communication interface. Once the reset of the controller is triggered, the traffic light system starts to work with the newly configured values. As illustrated in Figure~\ref{fig:TLC}, the system output is represented by the LED signals which is a 7-bit vector. } \exampleEnd



 
 %

\begin{algorithm}[htb]
\small
\caption{\textit{HIVE} Algorithm}
	\label{algo:HIVE}
    \begin{flushleft}
        
        \textbf{Inputs: } Scenarios: $T\{t_1,t_2,... ,t_n\}$, RTL: $R\{r_1,r_2,... ,r_n\}$, Drivers: $d$\\
        \textbf{Outputs: } Verification: $V\{v_1,v_2,... ,v_n\}$ 
    \end{flushleft}
	\begin{algorithmic}[1]
            \For{each $t$ in $T$} \Comment{Section~\ref{subsec:testCases}}
            \State $f\gets$ \textit{compile(d + t)}
            \State $F.append(f)$
            \For{Each $r$ in $R$} 
                \State Spec.append([t,behavior]) \Comment{Section~\ref{subsec:specification}}
            \EndFor
            \EndFor
            \State \textit{M} $\gets$ flattened($R$)\Comment{Section~\ref{subsec:sysmodel}}
            \For{each \textit{f} in $F$}
                \State trace $\gets$ $simulate(f +M)$
                \State $S\gets signalRanking(trace)$

                \State $a\gets extractArtifacts(S,R,M)$ \Comment{Section~\ref{subsec:extract}}
            \State $h\gets hintGeneration(S,M(a),\tau)$ \Comment{Section~\ref{subsec:Hints}}
		\State $H.append(h)$
            \EndFor
            \State I $\gets smt(R,d)$ \Comment{Section~\ref{subsec:proofs}}
            \For{each $r$ in $R$} 
                \For{each $t$ in $T$} 
                    \State V.append( $I(H_t^r)\subseteq Spec_t^r $) \Comment{ Pass/ Fail}
                \EndFor
            \EndFor
	      \State \textbf{Return} V
	\end{algorithmic} 
\end{algorithm}

\vspace{-0.1in}
\subsection{Test Cases for Verification Scenarios}\label{subsec:testCases}
In this step, we outline the functional scenarios that we want to provide guarantee with formal verification. For this purpose, we utilize the same test plan that is typically developed by designers to test the firmware. This may include unit tests for each of the components in the design under test, which are written in C or assembly (\textit{*.c/*.asm}) for testing the firmware. \textit{HIVE} can effectively use these unit tests to identify the behavior of the model related to the specific test scenarios. Note that one test case would be enough to state one test scenario. For example, if the firmware uses a 64-bit adder (one scenario), we only need to specify one testcase (e.g., adding 2 and 3 should produce 5) for verifying the addition scenario instead of $2^{128}$ testcases. These tests should cover all the scenarios that are expected to get formally verified.

After writing each of the test cases, we compile the firmware with device drivers using the relevant toolchain of the processor instruction-set architecture. Note that we keep only the test case as the main function of the SoC during this process. This will generate a separate firmware for each of the verification scenarios. Next, we need to simulate the design to obtain the behavioral model of the system for each scenario.

\vspace{0.05in}
\noindent \underline{{Example 2 (Test Cases)}}: 
\textit{Let us verify the functionality of the traffic light controller. We have to write test cases for the major verification scenarios. In this example, we have considered four simple test scenarios: SoC requesting sensor data from an external device (\textbf{Scenario 1}), SoC reading sensor data from the external device (\textbf{Scenario 2}), firmware requesting the traffic light controller for the side street green light (\textbf{Scenario 3}), and firmware requesting a pedestrian crossing instance for the main road (\textbf{Scenario 4}). These verification scenarios and corresponding test cases are derived from the test plan. Listing~\ref{code:case1} presents the test case written in the firmware in \textit{c} language for the first scenario. These test cases will be compiled separately with the device drivers in order to obtain four separate binary/hex files corresponding to each test scenario. } \exampleEnd

\begin{code}
    \caption{Firmware test case related to Scenario 1.} \label{code:case1}
    \vspace{-0.05in}
    \begin{center}
        \begin{minted}
[
% frame=lines,
% framesep=2mm,
% baselinestretch=1.2,
fontsize=\scriptsize,
linenos,
xleftmargin=2em,
]
{c}
#include <stddef.h>
#include "driver.h" 

int main(void) {
    char msg[] = "request()\r\n";
    uart_init(); // Initialize UART
    uart_write(msg, strlen(msg)); // Write msg to UART
    return 0;
}
\end{minted}
    \end{center}
    \vspace{-0.1in}
\end{code}
\subsection{Development of Specification}\label{subsec:specification}
Once we have each of the test scenarios, we need to capture the expected behavior of the system in a formal model. We write the functionality of each of the components with respect to each scenario. We put them together to obtain the formal model of the specification. 
\begin{code}
    \caption{Specification of UART module for Scenario 1.} \label{code:spec1}
    \vspace{-0.05in}
    \begin{center}
        \begin{minted}
[
% frame=lines,
% framesep=2mm,
% baselinestretch=1.2,
fontsize=\scriptsize,
linenos,
xleftmargin=2em,
]
{racket}
(define (send-byte byte)
  ; Initialize sending process
  (send-bit #f)
  ; sending the data as bits
  (for-range 0 8 (lambda (i) (send-bit 
        (bitvector->bool (extract i i byte)))))
  ; Ending the message
  (send-bit #t))
....
\end{minted}
    \end{center}
    \vspace{-0.1in}
\end{code}

\noindent \underline{{Example 3 (Specification)}}: 
\textit{Since Example 2 has four scenarios, we need to write four specifications for each of the module in the SoC (Example 1). Listing~\ref{code:spec1} shows a part of the specification that we have developed to capture the expected behavior of UART for the first scenario in the Racket programming syntax (*.rkt). In the example code, the \mintinline{racket}{send-byte} function initiates a data transmission process, sends each bit of an input byte sequentially, and concludes the transmission by signaling the end of the message. The actual bit transmission is delegated to an external function named \mintinline{racket}{send-bit} which describes the low level specification of the data transmission.} \exampleEnd
\vspace{-0.1in}

\subsection{Automated Generation of System Models}\label{subsec:sysmodel}

At this stage, we have compiled a binary for each of the verification scenarios, which is ready to be simulated with the implementation.  In order to proceed with the simulation of each scenario, we construct the system model. {\color{black} For this, we include the binary in the hardware description (e.g., Verliog allows reading compiled binary as a memory image). At this point, if we simulate the design, all the tasks that are implemented in the firmware will be simulated. This system model (firmware+hardware) file is then flattened to remove the design hierarchy. This enables equivalence checking (via symbolic simulation) of the entire implementation at the final stage with both hardware and firmware (driver/software).  It is important to keep the signal names preserved with the hierarchy during this stage since we need to align the dynamic simulation-related behavioral details with module-level architectural details. This process is repeated for all the firmware binaries that we have created in previous step. This results in multiple copies of system models with the same hardware, implementing each test scenario in the firmware separately. }

Next, we need to perform dynamic analysis. Dynamic analysis is performed by analyzing the trace data after performing the concrete simulations. For this purpose, we automatically generate the testbenches to simulate the SoC. This also includes the template to include external inputs to the system, if required for the particular test case. The objective of this step is to create a functional system model that passes the particular test case. 

\begin{figure}[H]
    \begin{center}
    \vspace{-0.15in}
    \begin{adjustwidth*}{-1em}{0.5em}
        \small
        \input{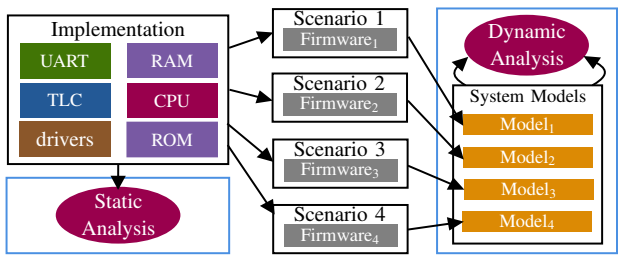}
    \end{adjustwidth*}
    \end{center}
      \vspace{-0.1in}
      \caption{System models generated for the example SoC. }
        \vspace{-0.1in}
      \label{fig:examplemodels}
\end{figure}

\vspace{0.05in}
\noindent \underline{{Example 4 (System Models)}}: \textit{For the example SoC, we have four verification scenarios in Example 2. As a result, we will have four separate system models which contain relevant test case firmware for the relevant scenario. Figure~\ref{fig:examplemodels} illustrates the individual system models for dynamic analysis (concrete simulation) while module-level implementation is utilized for static analysis.} \exampleEnd

\subsection{\textbf{Generation of Supporting Artifacts}}\label{subsec:extract}

We define proof-supporting artifacts as objects that can be derived from the system model. \textit{HIVE} utilizes both module-level static analysis and system-level dynamic analysis to extract supporting artifacts.  We first describe these two analysis techniques. Next, we discuss how to combine the extracted artifacts to infer meaningful details about the implementation. 

\subsubsection{\textbf{Module-Level Static Analysis}} 
\hfill\\
The purpose of the static analysis is to identify the structure of the implementation. 
This includes information such as signal dependencies, path conditions, and state register details. In order to identify these details, \textit{HIVE} extracts the abstract syntax tree (AST) of each module. AST contains information about  control flow and data flow related to the implementation of the module. Further, \textit{HIVE} extracts the Finite-State-Machine (FSM) details present in the module implementation.

\begin{figure}[htp]
\vspace{-0.1in}

        \begin{subfigure}{1\linewidth}
          \centering
          \input{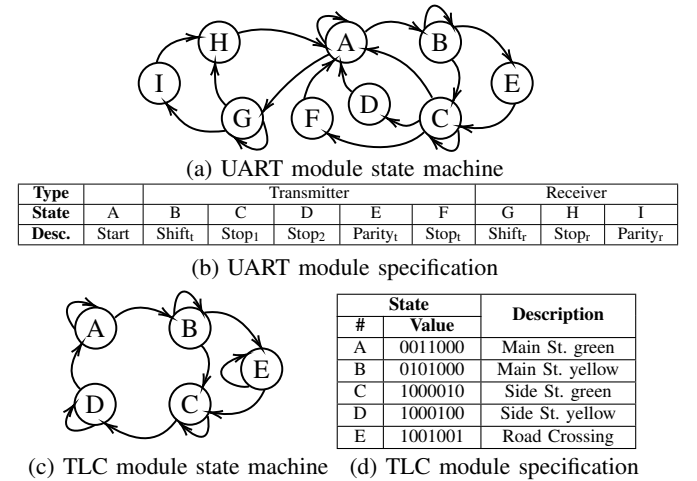}
          \vspace{-0.1in}
          \caption{UART module state machine}
          \label{fig:fsm_uart}
        \end{subfigure}%
        \hspace{3em}
        \begin{subfigure}{1\linewidth}
          \centering
        { \footnotesize
        \resizebox{\textwidth}{!}{
          \begin{tabular}{|c|c|ccccc|ccc|}
            \hline
            \textbf{Type} &  & \multicolumn{5}{c|}{{Transmitter}} & \multicolumn{3}{c|}{{Receiver}} \\ \hline
            \textbf{State} & {A} & \multicolumn{1}{c|}{{B}} & \multicolumn{1}{c|}{{C}} & \multicolumn{1}{c|}{{D}} & \multicolumn{1}{c|}{{E}} & {F} & \multicolumn{1}{c|}{{G}} & \multicolumn{1}{c|}{{H}} & {I} \\ \hline
            \textbf{Desc.} & Start & \multicolumn{1}{c|}{Shift\textsubscript{t}} & \multicolumn{1}{c|}{Stop\textsubscript{1}} & \multicolumn{1}{c|}{Stop\textsubscript{2}} & \multicolumn{1}{c|}{Parity\textsubscript{t}} & Stop\textsubscript{t} & \multicolumn{1}{c|}{Shift\textsubscript{r}} & \multicolumn{1}{c|}{Stop\textsubscript{r}} & Parity\textsubscript{r} \\ \hline
            \end{tabular}}
        }
          \vspace{-0.05in}
          \caption{UART module specification}
          \label{fig:spec_uart}
        \end{subfigure}



        \begin{subfigure}{.5\linewidth}
          \centering
          \tikzset{every picture/.style={line width=0.75pt}} 

\begin{tikzpicture}[x=0.75pt,y=0.75pt,yscale=-0.6,xscale=0.6]

\draw   (33.4,48.3) .. controls (33.4,38.75) and (41.14,31) .. (50.7,31) .. controls (60.25,31) and (68,38.75) .. (68,48.3) .. controls (68,57.86) and (60.25,65.6) .. (50.7,65.6) .. controls (41.14,65.6) and (33.4,57.86) .. (33.4,48.3) -- cycle ;

\draw   (112.65,48.3) .. controls (112.65,38.75) and (120.39,31) .. (129.95,31) .. controls (139.5,31) and (147.25,38.75) .. (147.25,48.3) .. controls (147.25,57.86) and (139.5,65.6) .. (129.95,65.6) .. controls (120.39,65.6) and (112.65,57.86) .. (112.65,48.3) -- cycle ;

\draw   (112.65,112.3) .. controls (112.65,102.75) and (120.39,95) .. (129.95,95) .. controls (139.5,95) and (147.25,102.75) .. (147.25,112.3) .. controls (147.25,121.86) and (139.5,129.6) .. (129.95,129.6) .. controls (120.39,129.6) and (112.65,121.86) .. (112.65,112.3) -- cycle ;

\draw   (32.9,112.48) .. controls (32.9,102.92) and (40.64,95.18) .. (50.2,95.18) .. controls (59.75,95.18) and (67.5,102.92) .. (67.5,112.48) .. controls (67.5,122.03) and (59.75,129.78) .. (50.2,129.78) .. controls (40.64,129.78) and (32.9,122.03) .. (32.9,112.48) -- cycle ;

\draw   (172.9,82.73) .. controls (172.9,73.17) and (180.64,65.43) .. (190.2,65.43) .. controls (199.75,65.43) and (207.5,73.17) .. (207.5,82.73) .. controls (207.5,92.28) and (199.75,100.03) .. (190.2,100.03) .. controls (180.64,100.03) and (172.9,92.28) .. (172.9,82.73) -- cycle ;

\draw    (66.88,39.6) .. controls (82.55,25.11) and (101.61,32.29) .. (112.69,38.4) ;
\draw [shift={(114.38,39.35)}, rotate = 210.17] [color={rgb, 255:red, 0; green, 0; blue, 0 }  ][line width=0.75]    (10.93,-3.29) .. controls (6.95,-1.4) and (3.31,-0.3) .. (0,0) .. controls (3.31,0.3) and (6.95,1.4) .. (10.93,3.29)   ;
\draw    (37.63,99.6) .. controls (26.27,88.49) and (29.8,75.9) .. (38.07,62.75) ;
\draw [shift={(39.13,61.1)}, rotate = 123.21] [color={rgb, 255:red, 0; green, 0; blue, 0 }  ][line width=0.75]    (10.93,-3.29) .. controls (6.95,-1.4) and (3.31,-0.3) .. (0,0) .. controls (3.31,0.3) and (6.95,1.4) .. (10.93,3.29)   ;
\draw    (119.13,124.85) .. controls (102.06,149.23) and (79.76,141.91) .. (59.65,129.25) ;
\draw [shift={(58.11,128.27)}, rotate = 32.88] [color={rgb, 255:red, 0; green, 0; blue, 0 }  ][line width=0.75]    (10.93,-3.29) .. controls (6.95,-1.4) and (3.31,-0.3) .. (0,0) .. controls (3.31,0.3) and (6.95,1.4) .. (10.93,3.29)   ;
\draw    (139.63,63.1) .. controls (150.55,74.26) and (144.61,87.58) .. (141.32,96.03) ;
\draw [shift={(140.63,97.85)}, rotate = 289.98] [color={rgb, 255:red, 0; green, 0; blue, 0 }  ][line width=0.75]    (10.93,-3.29) .. controls (6.95,-1.4) and (3.31,-0.3) .. (0,0) .. controls (3.31,0.3) and (6.95,1.4) .. (10.93,3.29)   ;
\draw    (142.13,35.35) .. controls (166.97,31.73) and (179.72,48.37) .. (189.18,63.76) ;
\draw [shift={(190.2,65.43)}, rotate = 238.82] [color={rgb, 255:red, 0; green, 0; blue, 0 }  ][line width=0.75]    (10.93,-3.29) .. controls (6.95,-1.4) and (3.31,-0.3) .. (0,0) .. controls (3.31,0.3) and (6.95,1.4) .. (10.93,3.29)   ;
\draw    (192.63,100.1) .. controls (183.9,120.23) and (167.17,122.71) .. (148.15,119.42) ;
\draw [shift={(146.38,119.1)}, rotate = 10.75] [color={rgb, 255:red, 0; green, 0; blue, 0 }  ][line width=0.75]    (10.93,-3.29) .. controls (6.95,-1.4) and (3.31,-0.3) .. (0,0) .. controls (3.31,0.3) and (6.95,1.4) .. (10.93,3.29)   ;
\draw    (33.38,49.35) .. controls (20.7,33.51) and (25.37,20.04) .. (48.86,29.25) ;
\draw [shift={(50.7,30)}, rotate = 203.01] [color={rgb, 255:red, 0; green, 0; blue, 0 }  ][line width=0.75]    (10.93,-3.29) .. controls (6.95,-1.4) and (3.31,-0.3) .. (0,0) .. controls (3.31,0.3) and (6.95,1.4) .. (10.93,3.29)   ;
\draw    (50.2,129.78) .. controls (29.9,140.33) and (17.74,132.99) .. (31.77,113.96) ;
\draw [shift={(32.9,112.48)}, rotate = 128.09] [color={rgb, 255:red, 0; green, 0; blue, 0 }  ][line width=0.75]    (10.93,-3.29) .. controls (6.95,-1.4) and (3.31,-0.3) .. (0,0) .. controls (3.31,0.3) and (6.95,1.4) .. (10.93,3.29)   ;
\draw    (116.63,36.85) .. controls (117.11,12.23) and (129.25,11.38) .. (139.35,33.13) ;
\draw [shift={(140.13,34.85)}, rotate = 246.43] [color={rgb, 255:red, 0; green, 0; blue, 0 }  ][line width=0.75]    (10.93,-3.29) .. controls (6.95,-1.4) and (3.31,-0.3) .. (0,0) .. controls (3.31,0.3) and (6.95,1.4) .. (10.93,3.29)   ;
\draw    (144.88,122.35) .. controls (147.8,140.15) and (140.98,147.01) .. (123.03,129.96) ;
\draw [shift={(121.63,128.6)}, rotate = 44.62] [color={rgb, 255:red, 0; green, 0; blue, 0 }  ][line width=0.75]    (10.93,-3.29) .. controls (6.95,-1.4) and (3.31,-0.3) .. (0,0) .. controls (3.31,0.3) and (6.95,1.4) .. (10.93,3.29)   ;
\draw    (179.63,97.35) .. controls (167.07,100.55) and (136.07,84.11) .. (175.52,71.19) ;
\draw [shift={(177.38,70.6)}, rotate = 162.9] [color={rgb, 255:red, 0; green, 0; blue, 0 }  ][line width=0.75]    (10.93,-3.29) .. controls (6.95,-1.4) and (3.31,-0.3) .. (0,0) .. controls (3.31,0.3) and (6.95,1.4) .. (10.93,3.29)   ;

\draw (129.95,112.3) node   [align=left] {\begin{minipage}[lt]{23.53pt}\setlength\topsep{0pt}
\begin{center}
C
\end{center}

\end{minipage}};
\draw (50.7,48.3) node   [align=left] {\begin{minipage}[lt]{23.53pt}\setlength\topsep{0pt}
\begin{center}
A
\end{center}

\end{minipage}};
\draw (129.95,48.3) node   [align=left] {\begin{minipage}[lt]{23.53pt}\setlength\topsep{0pt}
\begin{center}
B
\end{center}

\end{minipage}};
\draw (50.2,112.48) node   [align=left] {\begin{minipage}[lt]{23.53pt}\setlength\topsep{0pt}
\begin{center}
D
\end{center}

\end{minipage}};
\draw (190.2,82.73) node   [align=left] {\begin{minipage}[lt]{23.53pt}\setlength\topsep{0pt}
\begin{center}
E
\end{center}

\end{minipage}};

\end{tikzpicture}
          \vspace{-0.05in}
          \caption{TLC module state machine}
          \label{fig:fsm_tlc}
        \end{subfigure}%
        \hspace{-1em}
        \begin{subfigure}{.5\linewidth}
          \centering
        { \scriptsize
        \centering
          \begin{tabular}{|cc|c|}
            \hline
            \multicolumn{2}{|c|}{\textbf{State}} & \multirow{2}{*}{\textbf{Description}} \\ \cline{1-2}
            \multicolumn{1}{|c|}{\textbf{\#}} & \textbf{Value} &  \\ \hline
            \multicolumn{1}{|c|}{A} & 0011000 & Main St. green \\ \hline
            \multicolumn{1}{|c|}{B} & 0101000 & Main St. yellow \\ \hline
            \multicolumn{1}{|c|}{C} & 1000010 & Side St. green \\ \hline
            \multicolumn{1}{|c|}{D} & 1000100 & Side St. yellow \\ \hline
            \multicolumn{1}{|c|}{E} & 1001001 & Road Crossing \\ \hline
            \end{tabular}
        }
          \vspace{-0.05in}
          \caption{TLC module specification}
          \label{fig:spec_tlc}
        \end{subfigure}

      \caption{Finite State Machine (FSM) extracted from the UART and TLC modules by \textit{HIVE} and the relevant specifications.}
        \vspace{-0.1in}
      \label{fig:TLC_FSM}
\end{figure}

\vspace{0.05in}
\noindent \underline{{Example 5 (Static Analysis Artifacts)}}: \textit{For the SoC in Example 1, \textit{HIVE} was able to automatically extract all the module level ASTs along with two main FSMs related to the validation scenarios. Figure~\ref{fig:fsm_uart} shows the FSM extracted from the UART module while Figure~\ref{fig:fsm_tlc} shows the FSM extracted from the TLC module. Note that these figures show the visual representation of the FSMs. However, \textit{HIVE} internally represent them as comma delimited ASCII text file (\textit{*.kiss}) format. Figure~\ref{fig:spec_uart} and Figure~\ref{fig:spec_tlc} provides the details extracted from the specification to identify each state in the FSM. For example, $Stop_1$, $Stop_2$, and $Stop_t$ refer to transmitter communication halt states at 1/2 byte, byte, and final stop, respectively. Similarly, $Stop_r$ refers to the receiver's final stop. Likewise, $Shift_t$ and $Shift_r$ refer to transmitter and receiver shift states, respectively. Finally, $Parity_t$ and $Parity_r$ refer to transmitter and receiver parity check states, respectively. \exampleEnd}

\subsubsection{\textbf{System-Level Dynamic Analysis}}
\hfill\\
The purpose of the dynamic analysis is to automatically identify the behavioral model of the system for a specific scenario. \textit{HIVE} simulates the designs with the system models generated in Section~\ref{subsec:sysmodel} and obtains the simulation traces related to each scenario separately. Next,  the following procedure is applied to the simulation trace of each model separately.

Each of the signals in the model is analyzed and ranked according to their value change frequency. Algorithm~\ref{algo:ranking} illustrates the main steps involved in the signal ranking process. First, it extracts the signal list in the execution trace and generates a data structure to store the history related to each signal. Next, it analyses the simulation trace while appending the concrete value of each signal observed in the simulation, followed by sorting the final signal list based on the increasing order of value change frequency observed for each signal. Note that all the unknown logic (uninitialized) value states are considered for weakening and ranked last, while high-impedance states are taken into account for the signal ranking process. The ranking procedure serves as the heuristic for the hint generation process to identify the hint type corresponding to signals (\textit{i.e.}, if a certain signal is changed less than a certain threshold ($\tau$), consider for abstraction).

\begin{algorithm}[htb]
\small
\caption{signalRanking($vcd$)}
	\label{algo:ranking}
    \begin{flushleft}
        \textbf{Input:} Value Change Dump (VCD) Traces: $vcd$\\
        \textbf{Output:} Ranked Signals: $S$ 
    \end{flushleft}
	\begin{algorithmic}[1]
            \State \textit{Signals} $\gets$ extractSignals($vcd$)
            \For{each \textit{trace} in $vcd$}
                \State \{signal, value\} $\gets$ \textit{trace}
                \State \textit{Signals}$<$signal$>$.append(value)
            \EndFor
	    \State S $\gets$ \textbf{sort} \textit{Signals} based on \textbf{length}(value)
            
		\State Return $S$  
	\end{algorithmic} 
\end{algorithm}

\begin{figure}[htp]
    \begin{center}
    \vspace{-0.1in}

    

    



\footnotesize
\pgfplotsset{compat=newest}

\pgfplotsset{testbar/.style={
    xbar stacked,
    width=\columnwidth,
    xmajorgrids = true,
    xmin=0,xmax=100,
    ytick = data, 
    ytick style={draw=none},
    yticklabels = {Model\textsubscript{1},Model\textsubscript{2},Model\textsubscript{3},Model\textsubscript{4}},
    tick align = outside, xtick pos = left,
    bar width=4mm, y=5mm,
    nodes near coords={\scriptsize \pgfmathprintnumber{\pgfplotspointmeta}\%}, 
    xticklabel={\pgfmathprintnumber{\tick}\%},
    nodes near coords align={center}, 
    enlarge y limits=0.25, 
}}

\begin{tikzpicture}
    \begin{axis}[testbar,
            legend columns=4,
            legend style={
                    draw=none,
                    legend cell align=right,
                    at={(1.05,1.25)},
                    anchor=north east,
                    column sep=1ex
        }]
        \addplot[white ,fill=bblue] coordinates{(52,3) (54,2) (57,1) (58,0)};
        \addplot[white ,fill=rred] coordinates{(26,3) (33,2) (23,1) (26,0)};
        \addplot[white ,fill=ggreen] coordinates{(12,3) (7,2) (14,1) (11,0)};
        \addplot[white ,fill=ppurple] coordinates{(10,3) (6,2) (6,1) (5,0)};

        \addlegendimage{bblue,fill=bblue}
        \addlegendentry{Frequancy$=1$};
        \addlegendimage{rred,fill=rred}
        \addlegendentry{Frequancy$\leq 5$};
        \addlegendimage{ggreen,fill=ggreen}
        \addlegendentry{Frequancy$>5$};
        \addlegendimage{ppurple,fill=ppurple}
        \addlegendentry{unknown};

    \end{axis}
\end{tikzpicture}
    \end{center}
      \vspace{-0.1in}
      \caption{Signal ranking calculated by \textit{HIVE} on different test cases by analysis of simulation traces corresponding to each concrete simulation scenario. ($\tau=5$)}
      \label{fig:signal_ranking}
\end{figure}
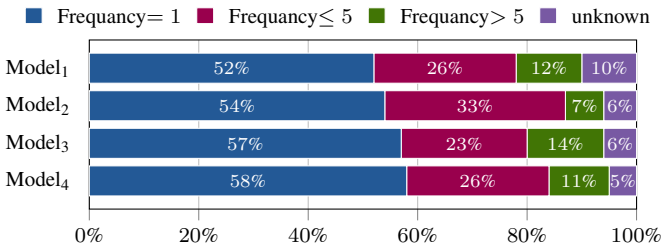

\noindent \underline{{Example 6 (Dynamic Analysis Artifacts)}}: \textit{After performing dynamic analysis on  models in Example 4, \textit{HIVE} obtains a data structure related to the signal history of each of the models and ranks the signals. Figure~\ref{fig:signal_ranking} illustrates the different percentages of signals for different frequencies of signal value changes found for each model separately. 
\exampleEnd}

\subsubsection{\textbf{Signal Alignment}}

After performing both static analysis and dynamic analysis, we need to combine the architectural model with the behavioral model of the design. \textit{HIVE} aligns the control flow graph (CFG) of the flattened netlist with each sub-module CFG by referring to the signal names. From CFG, it extracts the term dependency graph for each of the state variables in sub-modules.
Then from the sub-modules which contain FMSs, state register behavior is aligned with the FSM. Later this is used for path prioritization during formal proofs.

\vspace{0.05in}
\noindent \underline{{Example 7 (Signal Alignment)}}: \textit{Flattened implementation of Example 1 SoC contains all signal names with  hierarchy. If we consider the signal ``soc.uart.recv\_buf\_data'' of the flattened netlist, this represents the signal  ``recv\_buf\_data'' of the UART module. HIVE automatically decodes such information from the signal name and maps it with the corresponding module-level signal. 
\exampleEnd}

\subsection{Generation of System-level Hints}\label{subsec:Hints}

In \textit{HIVE}, hint generation is a fully automated process, where the user does not need to know  
implementation details about the design. Firmware is typically developed on top of hardware design created by third parties, and verification of these system environments is made possible by our proposed technique. In this section, we discuss the process of generating effective hints to simplify proofs by utilizing the proof-supporting artifacts derived in Section~\ref{subsec:extract}. This step of \textit{HIVE} serves two purposes. First, it automatically identifies each variable in the design to be used as a state or symbolic variable, and generates abstract syntax for each of the hints by analyzing the proof supporting artifacts. Next, it translates abstract syntax into a form that can be used by the formal verification framework.

Algorithm~\ref{algo:hintsynth} illustrates the main steps involved in the module-level hint generation process. First, it iterates through the modules available in the implementation of the SoC. If there is an FSM in the  module, it aligns the corresponding state register with the ranked signal list. This signal is analyzed against proof supporting artifacts to perform path prioritization (discussed in Section~\ref{subsubsec:pathprioritization}). 
All the other signals in the module are also aligned with the corresponding signal from the ranked signal list. If there are multiple instantiations of the same component, those will have multiple occurrences corresponding to each alignment.
Then the aligned signals are added to the hint list corresponding to the module and categorized based on the prospect hint type. First, it evaluates the signals whose value got changed only once during the concrete simulation. These are candidate signals to consider for concretization.  These signals are analyzed with the FSM states and path conditions. Specifically, a signal is appended for concretization if it got its value assignment in a state/path that is not in the current state/path conditions for the scenario under test.  

In \textit{HIVE}, the user has the ability to control the threshold ($\tau$) to consider a signal for overapproximation (default $\tau=5$, and this gets verified on the system model). All the uninitialized signals are considered for weakening. Since the back-end solver only has access to a single sub-problem at a time, the hint synthesis step performs solver queries on the system model in order to verify hints before committing them during the equivalence checking proofs.

\begin{algorithm}[htb]
\small
\caption{hintGeneration($S,M,\tau$)}
	\label{algo:hintsynth}
    \begin{flushleft}
        \textbf{Input: } RankedSignals $S$, Modules $M(s, fsm)$, Threshold. $\tau$\\
        \textbf{Output: } Hints $H$ 
    \end{flushleft}
	\begin{algorithmic}[1]
            \For{each \textit{m} in \textit{M}}
                \State stateVars $\gets$ S\{state(m$<fsm>$)\} \Comment{get FSM state register}
                \State vars $\gets$ S\{m$<$s$>$\} \Comment{Signal alignment} 
                \State $p \gets $ pathPrioritization($S,stateVars$)
                \State tempH$<$weaken$>$.\textit{append(p)}
            \EndFor
            \State tempH, H $\gets \emptyset$
            \State tempH$<$sRegs$>$.append(stateVars)
            \For{each \textit{v} in \textit{Vars}}
                \If{length(v.value)==1}
                    \If{\textit{assignment(v)} $\notin$ current path condition}
                    \State tempH$<$concrete$>$.\textit{append(v)}
                    \EndIf
                \ElsIf{length(v.value) $\geq\tau$}
                    \State tempH$<$overaprx$>$.\textit{append(v)}
                \ElsIf{length(v.value) $<\tau$}
                    \State $d \gets $ \textit{getDepends(v)}
                    \If{\textit{allowed(d)}}
                        \State tempH$<$abstract$>$.\textit{append(v)}
                    \EndIf
                \ElsIf{\textit{uninit(v.value)}}
                    \State tempH$<$weaken$>$.\textit{append(v)}
                \EndIf
            \EndFor

            \For{each \textit{h} in \textit{tempH}}
                \If{ \textit{verify(h)}} \Comment{Verify hints on the system model}
                    \State H.\textit{append(h)}
                \EndIf
            \EndFor

		\State Return $H$  
	\end{algorithmic} 
\end{algorithm}

\vspace{0.1in}
\subsubsection{\textbf{Elaboration of Hint Types}}\label{subsubsec:symbolicreduction}
\hfill\\
Performing modifications to the symbols involved in the formal proofs such that they have more constraints can reduce the state space of the verification problem. For that purpose, \textit{HIVE} performs the following types of symbol alterations:

 \noindent\textit{\underline{Concretization}}: The process of concretization involves replacing symbolic values with concrete values that represent specific values that the system or component can take during execution. This allows verification of the system or component using specific input values relevant for the scenario under test, rather than analyzing the behavior of the system or component under all possible input values. Since we have the system model and the hint list generated by Algorithm~\ref{algo:hintsynth}, we perform system-level queries from the solver to verify the variables that can be concretized before using them in module-level proofs.

 \noindent\textit{\underline{Weakening}}: Weakening a variable will effectively prune the formal model of the system by imposing constraints for that variable. This reduces the size of the symbolic execution tree and improves the efficiency of the symbolic execution by considering only the most relevant path to the specific scenario under consideration.

\noindent\textit{\underline{Overapproximation and Abstraction}}: The overapproximation process will  replace a symbolic variable that is propagated through the system with a new variable. This makes the  analysis less precise but more tractable. Abstraction will replace the variable with a symbolic term if the variable only depends on allowed
dependencies, which starts with the module inputs, and adds the symbol to the allowed dependencies.

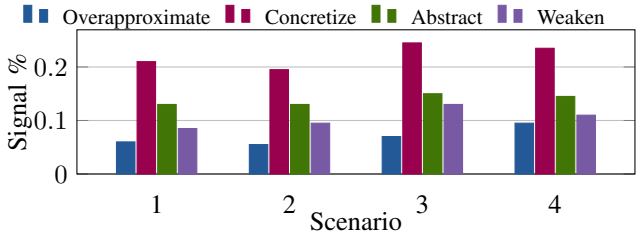
\begin{figure}[htp]
     \vspace{-0.1in}

    \begin{subfigure}{1\linewidth}
          \begin{adjustwidth*}{1em}{1em}
          \begin{tikzpicture}
\footnotesize
    \begin{axis}[
        ybar=2*\pgflinewidth,
        hide axis,
        xmin=10,
        xmax=50,
        ymin=0,
        ymax=0.4,
        legend columns=4,
        legend style={
                draw=none,
                legend cell align=left,
                column sep=1ex
        }
    ]



        \addlegendimage{bblue,fill=bblue}
        \addlegendentry{Overapproximate};
        \addlegendimage{rred,fill=rred}
        \addlegendentry{Concretize};
        \addlegendimage{ggreen,fill=ggreen}
        \addlegendentry{Abstract};
        \addlegendimage{ppurple,fill=ppurple}
        \addlegendentry{Weaken};

    \end{axis}
\end{tikzpicture}
          \end{adjustwidth*}
        \end{subfigure}%
        \vspace{-0.1in}
        \begin{subfigure}{1\linewidth}
          \centering





\begin{tikzpicture}
    \begin{axis}[
        width  = 9cm,
        height = 3.5cm,
        major x tick style = transparent,
        ybar=2*\pgflinewidth,
        bar width=7pt,
        ymajorgrids = true,
        ylabel = {Signal \%},
        xlabel = {Scenario},
        y label style={at={(axis description cs:-.14,0.5)},anchor=north},
        x label style={at={(axis description cs:.5,-0.2)},anchor=north},
        symbolic x coords={1,2,3,4},
        xtick = data,
        scaled y ticks = false,
        enlarge x limits=0.2,
        ymin=0,
        legend cell align=left,
        legend style={
                at={(0.5,1.05)},
                anchor=south east,
                column sep=1ex
        }
    ]
        \addplot[style={bblue,fill=bblue,mark=none}]
            coordinates {(1, 0.06) (2,0.055) (3, 0.07) (4,0.095)};

        \addplot[style={rred,fill=rred,mark=none}]
             coordinates {(1,0.21) (2,0.195) (3, 0.245) (4,0.235)};

        \addplot[style={ggreen,fill=ggreen,mark=none}]
             coordinates {(1,0.13) (2,0.13) (3, 0.15) (4,0.145)};

        \addplot[style={ppurple,fill=ppurple,mark=none}]
             coordinates {(1,0.085) (2,0.095) (3, 0.13) (4,0.11)};

    \end{axis}
\end{tikzpicture}
        \end{subfigure}%
   
      \vspace{-0.1in}
      \caption{Types of hints extracted and verified by \textit{HIVE} on different test scenarios and \% of hints on the system model.}
      \label{fig:signal_verify}
            \vspace{-0.1in}
\end{figure}

\vspace{0.05in}
\noindent \underline{{Example 8 (Generated Hints)}}: \textit{Figure~\ref{fig:signal_verify} illustrates the percentage of signals considered for different hint types for the four scenarios outlined in Example 2. After considering each of the signals as a hint for different modules, they are verified against the system model to check their validity.} \exampleEnd

\subsubsection{\textbf{Path Prioritization for Scenario Analysis}}\label{subsubsec:pathprioritization}
\begin{figure}[htp]
    \begin{center}
    \vspace{-0.1in}
        \tikzset{every picture/.style={line width=0.75pt}} 

\begin{tikzpicture}[x=0.75pt,y=0.75pt,yscale=-0.8,xscale=0.8]

\draw  [color={rgb, 255:red, 155; green, 155; blue, 155 }  ,draw opacity=1 ] (189.77,58.5) -- (194.46,48) -- (210.08,48) -- (214.77,58.5) -- (210.08,69) -- (194.46,69) -- cycle ;
\draw  [color={rgb, 255:red, 155; green, 155; blue, 155 }  ,draw opacity=1 ][dash pattern={on 0.84pt off 2.51pt}] (88.21,27) -- (92.9,16.5) -- (108.52,16.5) -- (113.21,27) -- (108.52,37.5) -- (92.9,37.5) -- cycle ;
\draw  [color={rgb, 255:red, 155; green, 155; blue, 155 }  ,draw opacity=1 ][dash pattern={on 0.84pt off 2.51pt}] (128.83,110.99) -- (133.52,100.49) -- (149.15,100.49) -- (153.83,110.99) -- (149.15,121.49) -- (133.52,121.49) -- cycle ;
\draw  [color={rgb, 255:red, 155; green, 155; blue, 155 }  ,draw opacity=1 ][dash pattern={on 0.84pt off 2.51pt}] (88.21,69) -- (92.9,58.5) -- (108.52,58.5) -- (113.21,69) -- (108.52,79.5) -- (92.9,79.5) -- cycle ;
\draw  [color={rgb, 255:red, 155; green, 155; blue, 155 }  ,draw opacity=1 ][dash pattern={on 0.84pt off 2.51pt}] (108.52,100.5) -- (113.21,90) -- (128.83,90) -- (133.52,100.5) -- (128.83,111) -- (113.21,111) -- cycle ;
\draw  [color={rgb, 255:red, 155; green, 155; blue, 155 }  ,draw opacity=1 ][dash pattern={on 0.84pt off 2.51pt}] (88.21,90) -- (92.9,79.5) -- (108.52,79.5) -- (113.21,90) -- (108.52,100.5) -- (92.9,100.5) -- cycle ;
\draw  [fill={rgb, 255:red, 153; green, 0; blue, 77 }  ,fill opacity=1 ] (128.83,69) -- (133.52,58.5) -- (149.15,58.5) -- (153.83,69) -- (149.15,79.5) -- (133.52,79.5) -- cycle ;
\draw  [color={rgb, 255:red, 74; green, 74; blue, 74 }  ,draw opacity=1 ] (149.15,58.5) -- (153.83,48) -- (169.46,48) -- (174.15,58.5) -- (169.46,69) -- (153.83,69) -- cycle ;
\draw   (128.83,90) -- (133.52,79.5) -- (149.15,79.5) -- (153.83,90) -- (149.15,100.49) -- (133.52,100.49) -- cycle ;
\draw  [color={rgb, 255:red, 0; green, 0; blue, 0 }  ,draw opacity=1 ][fill={rgb, 255:red, 35; green, 89; blue, 150 }  ,fill opacity=1 ] (128.83,48) -- (133.52,37.5) -- (149.15,37.5) -- (153.83,48) -- (149.15,58.5) -- (133.52,58.5) -- cycle ;
\draw  [color={rgb, 255:red, 0; green, 0; blue, 0 }  ,draw opacity=1 ] (108.52,58.5) -- (113.21,48) -- (128.83,48) -- (133.52,58.5) -- (128.83,69) -- (113.21,69) -- cycle ;
\draw  [color={rgb, 255:red, 0; green, 0; blue, 0 }  ,draw opacity=1 ] (108.52,79.5) -- (113.21,69) -- (128.83,69) -- (133.52,79.5) -- (128.83,90) -- (113.21,90) -- cycle ;
\draw   (149.15,100.49) -- (153.83,90) -- (169.46,90) -- (174.15,100.49) -- (169.46,110.99) -- (153.83,110.99) -- cycle ;
\draw  [color={rgb, 255:red, 128; green, 128; blue, 128 }  ,draw opacity=1 ] (169.46,69) -- (174.15,58.5) -- (189.77,58.5) -- (194.46,69) -- (189.77,79.5) -- (174.15,79.5) -- cycle ;
\draw  [color={rgb, 255:red, 0; green, 0; blue, 0 }  ,draw opacity=1 ][fill={rgb, 255:red, 35; green, 89; blue, 150 }  ,fill opacity=1 ] (169.46,48) -- (174.15,37.5) -- (189.77,37.5) -- (194.46,48) -- (189.77,58.5) -- (174.15,58.5) -- cycle ;
\draw  [fill={rgb, 255:red, 35; green, 89; blue, 150 }  ,fill opacity=1 ] (149.15,37.5) -- (153.83,27) -- (169.46,27) -- (174.15,37.5) -- (169.46,48) -- (153.83,48) -- cycle ;
\draw   (128.83,27) -- (133.52,16.5) -- (149.15,16.5) -- (153.83,27) -- (149.15,37.5) -- (133.52,37.5) -- cycle ;
\draw   (108.52,37.5) -- (113.21,27) -- (128.83,27) -- (133.52,37.5) -- (128.83,48) -- (113.21,48) -- cycle ;
\draw   (88.21,48) -- (92.9,37.5) -- (108.52,37.5) -- (113.21,48) -- (108.52,58.5) -- (92.9,58.5) -- cycle ;
\draw  [color={rgb, 255:red, 155; green, 155; blue, 155 }  ,draw opacity=1 ] (189.77,79.5) -- (194.46,69) -- (210.08,69) -- (214.77,79.5) -- (210.08,90) -- (194.46,90) -- cycle ;
\draw  [color={rgb, 255:red, 155; green, 155; blue, 155 }  ,draw opacity=1 ][dash pattern={on 0.84pt off 2.51pt}] (210.08,69) -- (214.77,58.5) -- (230.4,58.5) -- (235.08,69) -- (230.4,79.5) -- (214.77,79.5) -- cycle ;
\draw  [fill={rgb, 255:red, 35; green, 89; blue, 150 }  ,fill opacity=1 ] (230.4,58.5) -- (235.08,48) -- (250.71,48) -- (255.4,58.5) -- (250.71,69) -- (235.08,69) -- cycle ;
\draw   (230.4,79.5) -- (235.08,69) -- (250.71,69) -- (255.4,79.5) -- (250.71,90) -- (235.08,90) -- cycle ;
\draw  [color={rgb, 255:red, 155; green, 155; blue, 155 }  ,draw opacity=1 ][dash pattern={on 0.84pt off 2.51pt}] (210.08,90) -- (214.77,79.5) -- (230.4,79.5) -- (235.08,90) -- (230.4,100.49) -- (214.77,100.49) -- cycle ;
\draw  [color={rgb, 255:red, 0; green, 0; blue, 0 }  ,draw opacity=1 ][fill={rgb, 255:red, 220; green, 138; blue, 4 }  ,fill opacity=1 ] (189.77,100.49) -- (194.46,90) -- (210.08,90) -- (214.77,100.49) -- (210.08,110.99) -- (194.46,110.99) -- cycle ;
\draw  [fill={rgb, 255:red, 65; green, 117; blue, 5 }  ,fill opacity=1 ] (250.71,69) -- (255.4,58.5) -- (271.02,58.5) -- (275.71,69) -- (271.02,79.5) -- (255.4,79.5) -- cycle ;
\draw  [color={rgb, 255:red, 0; green, 0; blue, 0 }  ,draw opacity=1 ][fill={rgb, 255:red, 35; green, 89; blue, 150 }  ,fill opacity=1 ] (210.08,48) -- (214.77,37.5) -- (230.4,37.5) -- (235.08,48) -- (230.4,58.5) -- (214.77,58.5) -- cycle ;
\draw  [color={rgb, 255:red, 0; green, 0; blue, 0 }  ,draw opacity=1 ][fill={rgb, 255:red, 220; green, 138; blue, 4 }  ,fill opacity=1 ] (210.08,110.99) -- (214.77,100.49) -- (230.4,100.49) -- (235.08,110.99) -- (230.4,121.49) -- (214.77,121.49) -- cycle ;
\draw  [color={rgb, 255:red, 0; green, 0; blue, 0 }  ,draw opacity=1 ][fill={rgb, 255:red, 65; green, 117; blue, 5 }  ,fill opacity=1 ] (230.4,100.49) -- (235.08,90) -- (250.71,90) -- (255.4,100.49) -- (250.71,110.99) -- (235.08,110.99) -- cycle ;
\draw  [color={rgb, 255:red, 0; green, 0; blue, 0 }  ,draw opacity=1 ][fill={rgb, 255:red, 35; green, 89; blue, 150 }  ,fill opacity=1 ] (189.77,37.5) -- (194.46,27) -- (210.08,27) -- (214.77,37.5) -- (210.08,48) -- (194.46,48) -- cycle ;
\draw  [color={rgb, 255:red, 0; green, 0; blue, 0 }  ,draw opacity=1 ] (250.71,90) -- (255.4,79.5) -- (271.02,79.5) -- (275.71,90) -- (271.02,100.49) -- (255.4,100.49) -- cycle ;
\draw  [color={rgb, 255:red, 0; green, 0; blue, 0 }  ,draw opacity=1 ][fill={rgb, 255:red, 220; green, 138; blue, 4 }  ,fill opacity=1 ] (169.46,90) -- (174.15,79.5) -- (189.77,79.5) -- (194.46,90) -- (189.77,100.49) -- (174.15,100.49) -- cycle ;
\draw  [fill={rgb, 255:red, 220; green, 138; blue, 4 }  ,fill opacity=1 ] (149.15,79.5) -- (153.83,69) -- (169.46,69) -- (174.15,79.5) -- (169.46,90) -- (153.83,90) -- cycle ;
\draw    (142,85.1) .. controls (154.29,102.39) and (193.75,123.39) .. (215.5,128.6) .. controls (236.38,133.61) and (237.45,128.08) .. (245.02,116.13) ;
\draw [shift={(246,114.6)}, rotate = 123.18] [color={rgb, 255:red, 0; green, 0; blue, 0 }  ][line width=0.75]    (10.93,-3.29) .. controls (6.95,-1.4) and (3.31,-0.3) .. (0,0) .. controls (3.31,0.3) and (6.95,1.4) .. (10.93,3.29)   ;
\draw    (126,59.6) .. controls (112,45.1) and (133.5,25.1) .. (165,19.1) .. controls (196.5,13.1) and (210,13.1) .. (230,23.6) .. controls (249.3,33.73) and (252.3,38.28) .. (261.93,51.62) ;
\draw [shift={(263,53.1)}, rotate = 234.09] [color={rgb, 255:red, 0; green, 0; blue, 0 }  ][line width=0.75]    (10.93,-3.29) .. controls (6.95,-1.4) and (3.31,-0.3) .. (0,0) .. controls (3.31,0.3) and (6.95,1.4) .. (10.93,3.29)   ;
\draw  [draw opacity=0] (43,65) -- (74,65) -- (74,105) -- (43,105) -- cycle ;

\draw (141.33,69) node    {${\textstyle \textcolor[rgb]{1,1,1}{S}\textcolor[rgb]{1,1,1}{_{0}}}$};
\draw (263.21,69) node    {${\textstyle \textcolor[rgb]{1,1,1}{S}\textcolor[rgb]{1,1,1}{_{n}}}$};
\draw (174.5,132.2) node   [align=left] {\begin{minipage}[lt]{78.2pt}\setlength\topsep{0pt}
Scenario 1 Path
\end{minipage}};
\draw (289,18.5) node   [align=left] {\begin{minipage}[lt]{77.52pt}\setlength\topsep{0pt}
Scenario 2 Path
\end{minipage}};
\draw (242.9,100.49) node    {${\textstyle \textcolor[rgb]{1,1,1}{S}\textcolor[rgb]{1,1,1}{_{k}}}$};

\end{tikzpicture}
    \end{center}
      \vspace{-0.1in}
      \caption{Reduction of state space by execution path prioritization with respect to the scenarios outlined in firmware.}
        \vspace{-0.1in}
      \label{fig:paths}
\end{figure}
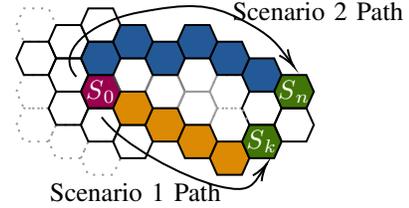

\hfill\\
The goal of path prioritization is  to focus the verification effort on the most relevant parts of the design. Most important paths are determined based on the test scenarios that are provided in the firmware application in Section~\ref{subsec:testCases}. During symbolic evaluation, these paths will get priority in the state space. Further, each scenario will be evaluated as a separate sub-problem. Scenario-based decomposition is particularly useful when dealing with complex control flow or data structures, where it may be difficult to analyze all possible behaviors in a single case. By dividing the problem into multiple scenarios, the verification effort can be focused on specific parts of the system, making the analysis more tractable. Figure~\ref{fig:paths} illustrates two instances where scenario-based decomposition is applied to a state register based on two test scenarios. 

Algorithm~\ref{algo:pathpriority} illustrates the major steps involved in the path prioritization process. First, we reconstruct the execution tree based on the test cases that the model was simulated on. Then traces that have different paths due to the test case difference are identified. For each case, we construct the hints such that path conditions for states that are not in the path get weakened. Path conditions are derived from the FSM that we extracted from the design during static analysis.

\begin{algorithm}[htb]
\small
\caption{pathPrioritization($S,S_r$)}
	\label{algo:pathpriority}
    \begin{flushleft}
        \textbf{Input:} Traces $S$, State Registers $S_r$,\\
        \textbf{Output:} Path Conditions $p$ 
    \end{flushleft}
	\begin{algorithmic}[1]
            \For{each v in \textit{S}}
    	    \If {v $ \in S_r$}
    		    \State \textit{v$<$exec$>$}.append(v.value) \Comment{Generate Execution Tree}
    		\EndIf
            \EndFor
            \For{each \textit{r} $\in$ $S_r$}
                \For{ each \textit{state} $\in$ \textit{r}}
                    \If {\textit{state} $\notin$ \textit{v$<$exec$>$}}
                        \State \textit{conditions} $\gets$ getConditions(\textit{State})
                        \State p.append(\textit{conditions})
                    \EndIf
                \EndFor
            \EndFor		
		\State Return $p$  
	\end{algorithmic} 
\end{algorithm}

\begin{figure}[htp]
    \begin{center}
        \begin{adjustwidth*}{-3.5em}{-3.5em}
            \input{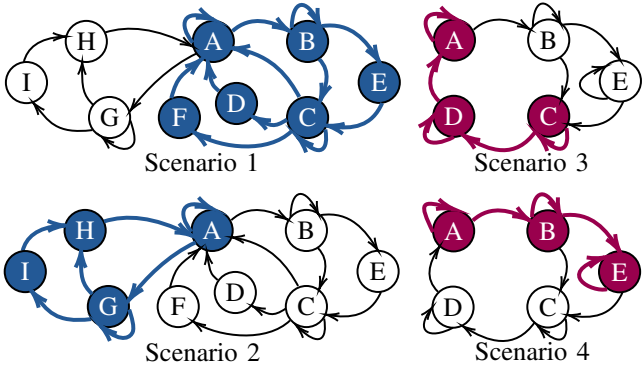}
        \end{adjustwidth*}
    \end{center}
      \vspace{-0.15in}
      \caption{Path prioritization for four scenarios outlined in Example 2 (colored paths get priority). }
      \label{fig:pathsforcases}
\end{figure}

%
\noindent \underline{{Example 9 (Path Prioritization) }}: 
\textit{Based on Algorithm~\ref{algo:pathpriority}, \textit{HIVE} is able to generate hints to guide the state space through relevant conditions of the example SoC. Figure~\ref{fig:pathsforcases} illustrates the FSM after applying the weakening conditions. For Scenario 1, path conditions related to the receiver (States H, I, and G) are weakened. For Scenario 2, path conditions related to the transmitter (States B, C, D, E, and F) are weakened. Similarly, in Scenario 3 and 4 irrelevant state conditions are weakened by the hints. In this case,  sample concretization hints that are related to the scenario 1  can be illustrated as follows. Let's assume that there is a variable $X$ that gets its values assigned within the states that are related to UART receiver (States H, I, and G). Since scenario 1 is not interested in states H, I, and G , the variable $X$ will be concretized to the value that was observed during the simulation. In this case, the SMT solver will process the variable $X$ as a concrete value instead of a  symbolic value.}\exampleEnd

\begin{figure*}[htp]
    \begin{center}
    \vspace{-0.2in}
            \input{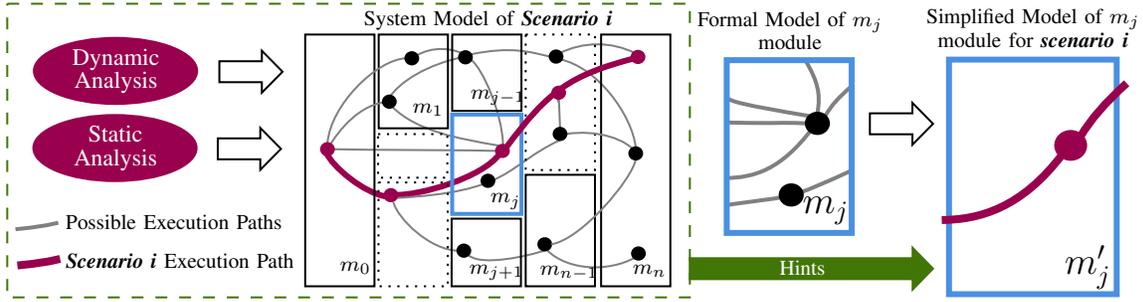}
    \end{center}
      \vspace{-0.1in}
      \caption{Illustration of how \textit{HIVE} simplify module level proofs utilizing system level hints.  Here $m_0-m_n$ represents different modules in the hardware implementation. In this instance, the system level hints simplify the formal model of the module $m_j$ for the \textit{\textbf{scenario i}} to $m'_j$ using constraints. Simplified $m'_j$ can be directly used in formal proof of \textit{\textbf{scenario i}}.}
        \vspace{-0.2in}
      \label{fig:decomposeSample}
\end{figure*}


\subsection{Scenario-based Decomposition}\label{subsec:subProblems}

Once hint generation is complete, \textit{HIVE} starts preparing for the hybrid symbolic execution. For this, natural boundaries of the designs such as module instantiations are used. Moreover, test scenarios are evaluated separately making them sub-problems of the original verification problem.
For ease of development and human understanding, system designs are composed of component (module)-level designs. Following the design hierarchy, system models can be decomposed into components. There are promising efforts for proof by decomposition with conjoining specifications and assume-guarantee reasoning~\cite{abadi1995conjoining,giannakopoulou2004assume}. The sub-problem decomposition technique of \textit{HIVE} is based on assume-guarantee reasoning.

In assume-guarantee reasoning, each subsystem or component is viewed as a black box, where the assumptions specify the possible behaviors of the environment or other components that interact with the subsystem, and the guarantees specify the expected behaviors of the subsystem under all possible assumptions. The technique then uses logical reasoning to prove that the guarantees of each subsystem are preserved in the presence of the assumptions of other subsystems or the environment.
In the paradigm of assume-guarantee, we define assumption ($A$) about the system, sub-component ($C$), and the property ($P$) that should hold within the system. The formula for the assume-guarantee relationship can be formulated as $f = \langle A \rangle \: C \langle P \rangle$. Here $f$ evaluates to true, whenever $C$ is a part of a system that satisfies $A$. In this case, the system must also guarantee $P$.

Consider a system $S$ that is composed of two sub-components of $C_1$ and $C_2$ ($S= C_1 || \: C_2$). In order to check whether the $S$ satisfies $P$, by analyzing $C_1$ and $C_2$  assume-guarantee reasoning principle in inference rule can be applied as follows. 

\begin{table}[H]
\centering
\small
\begin{tabular}{rl}
(Premise 1) & $\langle A \rangle \: C_1 \langle P \rangle$  \\
(Premise 2) & $\langle true \rangle \: C_2 \: \langle A \rangle$    \\ \hline
& $\langle true \rangle \: C_1 \: || \: C_2 \:\langle P \rangle$  
\end{tabular}
\vspace{-0.1in}
\end{table}

This indicates that if  $\langle A \rangle \: C_1 \langle P \rangle$ and $\langle true \rangle \: C_2 \langle A \rangle$ holds, then $\langle true \rangle \: S \:\langle P \rangle$ hold true. In order for the above inference rule to be valid, $A$ must be more abstract than $C_2$, and still it should describe $C_2$'s behavior correctly while $A$ should be strong enough for $C_1$ to satisfy $P$.

\textit{HIVE} make assumptions on the system model, validates them on the system model, and provides validated hints to the sub-component wise proofs. Therefore, \textit{HIVE} framework natively satisfies the definition of assume-guarantee.
Equivalence checking proofs can be performed on the decomposed system by selecting one component at a time with the hints generated by the \textit{HIVE}. Note that in situations that have the same set of hints in multiple cases, \textit{HIVE} will concatenate them to one sub-problem, and hence repeated analysis is avoided. Figure~\ref{fig:decomposeSample} illustrates a situation where \textit{HIVE} simplifies the formal model of the hardware module $m_j$ for \textit{scenario i} using the hints extracted from the system. Here, the dynamic analysis is performed with the simulation of \textit{scenario i}. \textit{HIVE} effectively removes unrelated components from the formal model using system-level hints, as discussed in Section~\ref{subsec:Hints}. The simplified model $m'_j$ is used in the verification of \textit{Scenario i}, as discussed in the next section.

\vspace{0.05in}
\noindent \underline{{Example 10 (Sub-Problems)}}: 
\textit{\textit{HIVE} is able to decompose the verification problem by module boundaries and verification scenarios. For the SoC in Example 1, we have provided four test cases. Based on both factors, \textit{HIVE} decomposes the entire system model into the following sub-problems}. 
\vspace{-0.05in}
\begin{samepage}
\begin{gather*}
     SoC_{scenario1} = \: cpu_1 \:||\: rom_1\: || \:ram_1\: || \:uart_1\: ||  \:tlc_1\:\\
     SoC_{scenario2} = \: cpu_2 \:||\: rom_2\: || \:ram_2\: || \:uart_2\: ||  \:tlc_2\:\\
     SoC_{scenario3} = \: cpu_3 \:||\: rom_3\: || \:ram_3\: || \:uart_3\: || \:tlc_3\: \\
     SoC_{scenario4} = \: cpu_4 \:||\: rom_4\: || \:ram_4\: || \:uart_4\: || \:tlc_4\: 
\end{gather*}\vspace{-0.2in}
\exampleEnd
\end{samepage}

\vspace{-0.1in}
\subsection{Sub-Problem Proofs using Hints}\label{subsec:proofs}

{So far we have discussed the process of automatically generating hints. In this section, we describe the steps to convert the implementation into a formal model and how the generated hints are utilized to address the path explosion. For this purpose, 
we convert a copy of the system model with inline firmware application (without test scenarios) to SMT format, which serves as our formal model for equivalence checking.} Then, the proposed framework will construct the state machine of the implementation against the state machine of the specification. Effective utilization of hints can simplify extremely complicated proof scenarios to tractable problems for the underlying SAT solvers. In this section, we discuss three types of limitations on scalable proof with symbolic execution. Then we discuss how hints are used in HIVE to mitigate these limitations. 
There are three major factors that affect the scalability of formal proofs: path explosion, state space explosion, and exhaustive components in the design.

\vspace{0.05in}
\noindent \textit{\underline{Path Explosion:} } Path explosion occurs in formal methods, such as model checking and symbolic execution, where the number of paths in a system's state space can grow exponentially with the size of the system. In other words, it becomes computationally infeasible to exhaustively check all possible paths for correctness or to explore the entire state space. As a result, the verification process may introduce unacceptable time and memory overhead, and in reality, may unsuccessfully terminate after exceeding memory capacity of the computer. \textit{HIVE} addresses path explosion by performing path prioritization relevant to the test scenarios outlined in Algorithm 4.

\vspace{0.05in}
\noindent \textit{\underline{Symbolic Term Explosion:} } Another problem faced by formal proofs is that verification techniques that rely on symbolic execution can lead to a symbolic term explosion problem, which occurs when the number of possible program states (combinations of symbolic and concrete values) grow too large to be handled by the analysis tool. This can occur when the symbolic execution engine explores all possible states through the program, which can be exponential in the size of the input space. \textit{HIVE} implements constraints on the symbolic variables in the form of concretization, weaken, overapproximation, and abstraction as outlined in Section~\ref{subsubsec:symbolicreduction}. This process has the ability to control the term space from symbolic to concrete while making some variables opaque so that their values can be substituted later in the process.

\vspace{0.05in}
\noindent \textit{\underline{Exhaustive Components:} } There are specific components in the design where a large run time can be anticipated during the proofs. Components such as memories and counters can contribute to the exponential state space expansion. This problem is also addressed by the path prioritization technique since all the unused areas of memory components will get lower priority by effectively abstracting such implementations to inputs and outputs with only relevant logic.

\vspace{0.05in}
Note that \textit{HIVE} generates the hint only for the scenario under test. Therefore, assumptions and properties are only valid for the specific scenario under consideration. 

\begin{code}
\caption{Sample set of overapproximate hints generated by the \textit{HIVE} and import procedure to utilize hints in the proofs.}  \label{code:samplehints-over}
    \begin{center}
    \vspace{-0.05in}
        \begin{minted}
[
% frame=lines,
% framesep=2mm,
% baselinestretch=1.2,
fontsize=\scriptsize,
linenos,
xleftmargin=2em,
]
{racket}
(define (overapproximate-case1 module)
    (match module
        [`uart (for ([field (list 
                    "recv_buf_valid"
                    "recv_buf_data"
                    "recv_divcnt"
                    "recv_pattern"
                    ... )])
                (match-abstract! field field))]
        [`cpu (for ([field (list 
                    ... )])
                (match-abstract! field field))]
        ....
        [default (displayln "Module not found!")]))
; Usage of hints in the proofs
(require "symcrete/dut/hints.rkt") ; import hints
(hint overapproximate-case1 `uart) ; use for UART proof
\end{minted}
    \end{center}
\end{code}

%
\noindent \underline{{Example 11 (Hints in \textit{Racket} Syntax) }}: 
\textit{For four testing scenarios in Example 2, \textit{HIVE} categorized signals into groups to be considered as hints in each scenario. Then these signals are adopted into pre-built \textit{Racket} programming language templates. A sample set of overapproximate hints after verifying them on the system model and adopting them to \textit{Racket} format are illustrated in Listing~\ref{code:samplehints-over}. 
Hints related to each case are constructed separately and generated hints can be directly imported into the proofs. The example  code defines a function \mintinline{racket}{overapproximate-case1} to generate hints for proofs related to scenario 1 of the  system model. It uses pattern matching to handle different sub-problems related to each of the module (e.g., uart and cpu), iterating over their respective fields and employing the  \mintinline{racket}{match-abstract!} function.
\exampleEnd}

\section{Experiments} \label{sec:experiments}

This section presents {four}  case studies using hardware and firmware implementations to demonstrate the effectiveness of \textit{HIVE}. First, we introduce the setup we have used to perform the experiments. Next, we discuss each case study and present results on hardware-firmware co-verification.

\subsection{Experimental Setup}
We consider {four}  System-on-Chip (SoC) implementations. These implementations utilize \textit{PicoRV32} processor~\cite{picorv} which is a CPU based on RISC-V instruction-set architecture configured with a ROM and RAM.  

To perform symbolic execution and equivalence checking, we have used \textit{Rosette}~\cite{torlak2013growing} language with \textit{Z3}~\cite{z3} as the back-end solver. Rosette is a solver-aided programming language extension for the \textit{Racket} functional programming language. In order to compile the \textit{C} and \textit{assembly} programs into binary, \textit{riscv-gnu-toolchain}~\cite{GCC} was used. To convert Verilog RTL implementations into \textit{smt2} (\textit{Satisfiability Modulo Theories Library v2}) format, \textit{Yosys}~\cite{Yosys} open synthesis tool was used. \textit{Rosette} package \textit{rtlv}~\cite{morozertlv} was used for lifting the smt2 output to racket supporting syntax and \textit{rtlv/shiva} was used to feed hints to the equivalence checking problems. Specifications related to each of the test scenarios are manually written in the \textit{Racket} language. Extraction of abstract syntax tree is performed by \textit{Icarus Verilog Loadable Target API}~\cite{icarusverilogAPI}. Module level FSM extraction and  data flow (\textit{*.blif}) are extracted through \textit{Yosys}. All the simulations for obtaining Value Change Dumps (VCD) were performed with Icarus Verilog~\cite{icarusverilog}. All the scripts related to hint generation, path prioritization, and symbolic state reduction were implemented in \textit{Python}. All the experiments were performed on a \textit{ARMv8.5} based \textit{Apple~M1} system with 16GB RAM inside a \textit{Docker} environment. 

For the evaluation of performance, we have selected {four} benchmark circuits that contain considerably large FSM implementations. Table~\ref{tab:socsizes} presents the size and trace length-related details about the benchmark SoCs. After the validation process, each of the SoC was compiled with \textit{Yosys} and \textit{nextpnr}~\cite{nextpnr} to generate the necessary binary file to program the \textit{Lattice ICE-Sugar-Nano} open-source Field Programmable Gate Array (FPGA). Finally, the usability of validated real-world SoCs were evaluated by running the firmware on the programmed FPGA.


\begin{table}[htp]
\centering
\small
\caption{Size comparison of RISC-V based SoC implementations considered for the {four} case studies.}\label{tab:socsizes} 
\footnotesize
\begin{tabular}{|l|r|r|r|r|}
\hline
Case Study SoC & \multicolumn{1}{c|}{\textbf{SEB}} & \multicolumn{1}{c|}{\textbf{ECC}} & \multicolumn{1}{c|}{\textbf{LCD}} & \multicolumn{1}{c|}{\textbf{SATA}} \\ \hline
Lines of Code (HW+FW) & 4371 & 4999 & 5022 &  5628\\ \hline
Number of Signals (HW) & 1471 & 1563 & 1494 & 1645 \\ \hline
Average Trace Size (MB) & 695 & 1874 & 1084 & 2027 \\ \hline
Simulated Clock Cycles & 50,000 & 100,000 & 100,000 & 50, 000 \\ \hline
\end{tabular}
\vspace{-0.1in}
\end{table}

\subsection{Case Study 1: Simple Encrypted Backup (SEB)}\label{subsec:dataSafe}

\begin{figure}[htp]
    \begin{center}
    \vspace{-0.1in}
        \tikzset{every picture/.style={line width=0.75pt}} 

\begin{tikzpicture}[x=0.75pt,y=0.75pt,yscale=-0.8,xscale=0.9]

\draw    (217.5,35.49) -- (238.63,35.6) ;
\draw    (217.5,15.47) -- (239.27,15.47) ;
\draw  [color={rgb, 255:red, 74; green, 144; blue, 226 }  ,draw opacity=1 ][line width=0.75]  (238.5,48.1) -- (342.83,48.1) -- (342.83,97.18) -- (238.5,97.18) -- cycle ;
\draw  [draw opacity=0][fill={rgb, 255:red, 93; green, 90; blue, 90 }  ,fill opacity=1 ] (238,3.02) -- (342.67,3.02) -- (342.67,20.82) -- (238,20.82) -- cycle ;

\draw    (132.17,77.45) -- (132.57,15.89) -- (140.86,15.89) ;
\draw    (123.86,41.15) -- (143.25,41.16) ;
\draw    (123.5,77.34) -- (141.5,77.4) ;
\draw  [draw opacity=0][fill={rgb, 255:red, 153; green, 0; blue, 77 }  ,fill opacity=1 ][line width=0.75]  (46.6,6.74) -- (125.5,6.74) -- (125.5,64.01) -- (46.6,64.01) -- cycle ;

\draw  [draw opacity=0][fill={rgb, 255:red, 120; green, 89; blue, 163 }  ,fill opacity=1 ][line width=0.75]  (139.85,55.51) -- (217.5,55.51) -- (217.5,89.02) -- (139.85,89.02) -- cycle ;
\draw  [draw opacity=0][fill={rgb, 255:red, 128; green, 128; blue, 128 }  ,fill opacity=1 ][line width=0.75]  (146.43,70.11) -- (210.04,70.11) -- (210.04,85.62) -- (146.43,85.62) -- cycle ;

\draw  [draw opacity=0][fill={rgb, 255:red, 120; green, 89; blue, 163 }  ,fill opacity=1 ][line width=0.75]  (46.79,67.99) -- (124.5,67.99) -- (124.5,89.09) -- (46.79,89.09) -- cycle ;

\draw  [draw opacity=0][fill={rgb, 255:red, 65; green, 117; blue, 5 }  ,fill opacity=1 ][line width=0.75]  (139.1,29.72) -- (217.5,29.72) -- (217.5,50.84) -- (139.1,50.84) -- cycle ;

\draw   (38,2.41) -- (226.33,2.41) -- (226.33,97.18) -- (38,97.18) -- cycle ;
\draw  [draw opacity=0][fill={rgb, 255:red, 35; green, 89; blue, 150 }  ,fill opacity=1 ][line width=0.75]  (140.57,6.87) -- (217.11,6.87) -- (217.11,23.91) -- (140.57,23.91) -- cycle ;

\draw  [draw opacity=0][fill={rgb, 255:red, 139; green, 87; blue, 42 }  ,fill opacity=1 ][line width=0.75]  (244,53.1) -- (288.5,53.1) -- (288.5,91.79) -- (244,91.79) -- cycle ;

\draw  [draw opacity=0][fill={rgb, 255:red, 220; green, 138; blue, 4 }  ,fill opacity=1 ][line width=0.75]  (292.5,53.1) -- (337,53.1) -- (337,91.8) -- (292.5,91.8) -- cycle ;

\draw [color={rgb, 255:red, 74; green, 144; blue, 226 }  ,draw opacity=1 ][line width=0.75]    (210.04,85.62) -- (237.5,97.18) ;
\draw [color={rgb, 255:red, 74; green, 144; blue, 226 }  ,draw opacity=1 ][line width=0.75]    (210.04,70.11) -- (238.5,48.1) ;
\draw  [draw opacity=0][fill={rgb, 255:red, 93; green, 90; blue, 90 }  ,fill opacity=1 ] (238.17,25.07) -- (342.83,25.07) -- (342.83,42.87) -- (238.17,42.87) -- cycle ;

\draw (178.84,15.44) node   [align=left] {\begin{minipage}[lt]{52.05pt}\setlength\topsep{0pt}
\begin{center}
\textcolor[rgb]{1,1,1}{{\footnotesize SPI}}
\end{center}

\end{minipage}};
\draw (178.23,77.91) node   [align=left] {\begin{minipage}[lt]{43.26pt}\setlength\topsep{0pt}
\begin{center}
\textcolor[rgb]{1,1,1}{{\footnotesize Firmware}}
\end{center}

\end{minipage}};
\draw (178.3,40.35) node   [align=left] {\begin{minipage}[lt]{53.31pt}\setlength\topsep{0pt}
\begin{center}
\textcolor[rgb]{1,1,1}{{\footnotesize UART}}
\end{center}

\end{minipage}};
\draw (85.64,78.61) node   [align=left] {\begin{minipage}[lt]{52.85pt}\setlength\topsep{0pt}
\begin{center}
\textcolor[rgb]{1,1,1}{{\footnotesize RAM}}
\end{center}

\end{minipage}};
\draw (86.05,35.56) node   [align=left] {\begin{minipage}[lt]{53.65pt}\setlength\topsep{0pt}
\begin{center}
\textcolor[rgb]{1,1,1}{{\footnotesize CPU}}
\end{center}

\end{minipage}};
\draw (314.75,72.57) node   [align=left] {\begin{minipage}[lt]{30.26pt}\setlength\topsep{0pt}
\begin{center}
\textcolor[rgb]{1,1,1}{{\footnotesize apps:}}\\\textcolor[rgb]{1,1,1}{{\scriptsize cipher()}}
\end{center}

\end{minipage}};
\draw (290.16,12.45) node   [align=left] {\begin{minipage}[lt]{70.94pt}\setlength\topsep{0pt}
\begin{center}
\textcolor[rgb]{1,1,1}{{\footnotesize External Flash}}
\end{center}

\end{minipage}};
\draw (266.25,60.94) node   [align=left] {\begin{minipage}[lt]{30.26pt}\setlength\topsep{0pt}
\begin{center}
\textcolor[rgb]{1,1,1}{{\footnotesize drivers:}}
\end{center}

\end{minipage}};
\draw (266.25,71.65) node   [align=left] {\begin{minipage}[lt]{30.26pt}\setlength\topsep{0pt}
\begin{center}
\textcolor[rgb]{1,1,1}{{\scriptsize uart()}}
\end{center}

\end{minipage}};
\draw (266.25,81.86) node   [align=left] {\begin{minipage}[lt]{30.26pt}\setlength\topsep{0pt}
\begin{center}
\textcolor[rgb]{1,1,1}{{\scriptsize flash()}}
\end{center}

\end{minipage}};
\draw (290.33,34.5) node   [align=left] {\begin{minipage}[lt]{70.94pt}\setlength\topsep{0pt}
\begin{center}
\textcolor[rgb]{1,1,1}{{\footnotesize Host Device}}
\end{center}

\end{minipage}};
\draw (178.75,62.62) node   [align=left] {\begin{minipage}[lt]{52.7pt}\setlength\topsep{0pt}
\begin{center}
\textcolor[rgb]{1,1,1}{{\footnotesize ROM}}
\end{center}

\end{minipage}};

\end{tikzpicture}
    \end{center}
      \vspace{-0.1in}
      \caption{Simple Encrypted Backup SoC.}
        \vspace{-0.1in}
      \label{fig:encSoC}
\end{figure}
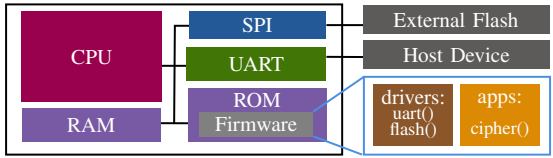

Simple Encrypted Backup (SEB) is a data logger SoC that reads data with a secret numeric pin from a host device and stores the result after performing a simple transposition cipher on the data on an external memory card. It uses the UART interface to communicate with the host device and uses an external memory card reader that communicates through the Serial Peripheral Interface (SPI) to store data as illustrated in Figure~\ref{fig:encSoC}. 
The external device can request to read the data from the storage by providing the secret pin back. External devices access the SoC via the terminal environment as illustrated in Figure~\ref{fig:terminalSoC}.

Both UART and SPI are configured on the SoC such that they both have separate MMIO regions to communicate with the CPU. The transposition cipher logic is implemented within the firmware. In order to evaluate the correctness of the encrypted backup SoC, we have developed a test plan with several test scenarios. This includes verifying reading from the host machine, writing to the host machine, reading from the external memory card, writing to the external memory card, encrypting the data, and decrypting the data. First, we attempted to perform equivalence checking on the system for outlined properties without \textit{HIVE}, and Z3 solver timeouts were observed for all scenarios. Therefore, we applied the proposed approach and validated the implementation. 

\begin{figure}[htp]

    \begin{center}
        \input{Images/terminalSoC}
    \end{center}
      \vspace{-0.2in}
      \caption{Simple Encrypted Backup terminal interface on a host machine, SoC is executed on the \textit{Ice-Sugar-Nano} \small{FPGA}.}
      \label{fig:terminalSoC}
\end{figure}

Table~\ref{tab:SEBresults} presents the results related to the verification process of the SEB SoC. The verification results illustrate that \textit{HIVE} is able to decompose the SoC verification problem into tractable sub-problems for the underlying solver within an acceptable time and memory.

\begin{table}[htp]
\footnotesize
\caption{Time (in Minutes) and memory (in Megabytes) consumed by the validation scenarios of the SEB  SoC. \textit{We did not show comparison with state-of-the-art since existing methods without hints face timeouts for all scenarios.}}\label{tab:SEBresults}
\centering
\vspace{-0.05in}
\resizebox{\columnwidth}{!}{
\begin{tabular}{|l|l|cc|cc|}
\hline
\multicolumn{1}{|c|}{\multirow{2}{*}{\textbf{Property}}} & \multicolumn{1}{c|}{\multirow{2}{*}{\textbf{Test Case}}} & \multicolumn{2}{c|}{\textbf{\begin{tabular}[c]{@{}c@{}}Hint\\ Generation\end{tabular}}} & \multicolumn{2}{c|}{\textbf{\begin{tabular}[c]{@{}c@{}}Equivalence\\ Checking\end{tabular}}} \\ \cline{3-6} 
\multicolumn{1}{|c|}{} & \multicolumn{1}{c|}{} & \multicolumn{1}{c|}{\textbf{Time}} & \multicolumn{1}{c|}{\textbf{Mem}} & \multicolumn{1}{c|}{\textbf{Time}} & \multicolumn{1}{c|}{\textbf{Mem}} \\ \hline
\multirow{2}{*}{Read} & \textit{from SPI Memory} & \multicolumn{1}{c|}{13} & 674 & \multicolumn{1}{c|}{19} & 639 \\ \cline{2-6} 
 & \textit{from Host} & \multicolumn{1}{c|}{17} & 756 & \multicolumn{1}{c|}{25} & 703 \\ \hline
\multirow{2}{*}{Write} & \textit{to SPI Memory} & \multicolumn{1}{c|}{15} & 698 & \multicolumn{1}{c|}{20} & 655 \\ \cline{2-6} 
 & \textit{to host} & \multicolumn{1}{c|}{18} & 789 & \multicolumn{1}{c|}{24} & 697 \\ \hline
\multirow{2}{*}{Text} & \textit{Encrypt} & \multicolumn{1}{c|}{21} & 701 & \multicolumn{1}{c|}{30} & 985 \\ \cline{2-6} 
 & \textit{Decrypt} & \multicolumn{1}{c|}{21} & 701 & \multicolumn{1}{c|}{32} & 946 \\ \hline
\end{tabular}}
\end{table}

\vspace{-0.1in}
\subsection{Case Study 2: ECC Core Accelerator}\label{subsec:ECCCore}

\begin{figure}[htp]
    \begin{center}
    \vspace{-0.1in}
        \tikzset{every picture/.style={line width=0.75pt}} 

\begin{tikzpicture}[x=0.75pt,y=0.75pt,yscale=-0.8,xscale=0.9]

\draw  [color={rgb, 255:red, 74; green, 144; blue, 226 }  ,draw opacity=1 ][line width=0.75]  (238.5,18.14) -- (342.83,18.14) -- (342.83,81.18) -- (238.5,81.18) -- cycle ;
\draw    (132.17,77.45) -- (132.57,15.89) -- (140.86,15.89) ;
\draw    (120.66,48.35) -- (140.05,48.36) ;
\draw    (132.17,77.45) -- (141.5,77.4) ;
\draw  [draw opacity=0][fill={rgb, 255:red, 153; green, 0; blue, 77 }  ,fill opacity=1 ][line width=0.75]  (46.6,29.74) -- (125.5,29.74) -- (125.5,87.01) -- (46.6,87.01) -- cycle ;

\draw  [draw opacity=0][fill={rgb, 255:red, 120; green, 89; blue, 163 }  ,fill opacity=1 ][line width=0.75]  (139.85,29.51) -- (217.5,29.51) -- (217.5,63.02) -- (139.85,63.02) -- cycle ;
\draw  [draw opacity=0][fill={rgb, 255:red, 120; green, 89; blue, 163 }  ,fill opacity=1 ][line width=0.75]  (139.79,67.66) -- (217.5,67.66) -- (217.5,88.77) -- (139.79,88.77) -- cycle ;

\draw   (38,2.41) -- (226.33,2.41) -- (226.33,97.18) -- (38,97.18) -- cycle ;
\draw  [draw opacity=0][fill={rgb, 255:red, 128; green, 128; blue, 128 }  ,fill opacity=1 ][line width=0.75]  (146.43,44.11) -- (210.04,44.11) -- (210.04,59.62) -- (146.43,59.62) -- cycle ;

\draw  [draw opacity=0][fill={rgb, 255:red, 35; green, 89; blue, 150 }  ,fill opacity=1 ][line width=0.75]  (47.8,6.87) -- (217.11,6.87) -- (217.11,23.91) -- (47.8,23.91) -- cycle ;

\draw  [draw opacity=0][fill={rgb, 255:red, 139; green, 87; blue, 42 }  ,fill opacity=1 ][line width=0.75]  (244,24.2) -- (288.5,24.2) -- (288.5,72.79) -- (244,72.79) -- cycle ;
\draw  [draw opacity=0][fill={rgb, 255:red, 220; green, 138; blue, 4 }  ,fill opacity=1 ][line width=0.75]  (292.5,24.2) -- (337,24.2) -- (337,72.8) -- (292.5,72.8) -- cycle ;

\draw [color={rgb, 255:red, 74; green, 144; blue, 226 }  ,draw opacity=1 ][line width=0.75]    (210.04,59.62) -- (238.5,81.18) ;
\draw [color={rgb, 255:red, 74; green, 144; blue, 226 }  ,draw opacity=1 ][line width=0.75]    (210.04,44.11) -- (238.5,18.14) ;

\draw (178.75,36.62) node   [align=left] {\begin{minipage}[lt]{52.7pt}\setlength\topsep{0pt}
\begin{center}
\textcolor[rgb]{1,1,1}{{\footnotesize ROM}}
\end{center}

\end{minipage}};
\draw (132.46,15.44) node   [align=left] {\begin{minipage}[lt]{115.13pt}\setlength\topsep{0pt}
\begin{center}
\textcolor[rgb]{1,1,1}{{\footnotesize ECC Core}}
\end{center}

\end{minipage}};
\draw (178.23,51.91) node   [align=left] {\begin{minipage}[lt]{43.26pt}\setlength\topsep{0pt}
\begin{center}
\textcolor[rgb]{1,1,1}{{\footnotesize Firmware}}
\end{center}

\end{minipage}};
\draw (178.64,78.28) node   [align=left] {\begin{minipage}[lt]{52.85pt}\setlength\topsep{0pt}
\begin{center}
\textcolor[rgb]{1,1,1}{{\footnotesize RAM}}
\end{center}

\end{minipage}};
\draw (86.05,58.56) node   [align=left] {\begin{minipage}[lt]{53.65pt}\setlength\topsep{0pt}
\begin{center}
\textcolor[rgb]{1,1,1}{{\footnotesize CPU}}
\end{center}

\end{minipage}};
\draw (314.75,48.66) node   [align=left] {\begin{minipage}[lt]{30.26pt}\setlength\topsep{0pt}
\begin{center}
\textcolor[rgb]{1,1,1}{{\footnotesize apps:}}\\\textcolor[rgb]{1,1,1}{{\scriptsize ecdsa()}}
\end{center}

\end{minipage}};
\draw (265.92,36.04) node   [align=left] {\begin{minipage}[lt]{30.26pt}\setlength\topsep{0pt}
\begin{center}
\textcolor[rgb]{1,1,1}{{\footnotesize drivers:}}
\end{center}

\end{minipage}};
\draw (265.92,48.49) node   [align=left] {\begin{minipage}[lt]{30.26pt}\setlength\topsep{0pt}
\begin{center}
\textcolor[rgb]{1,1,1}{{\scriptsize ecc{\tiny\_}arith()}}
\end{center}

\end{minipage}};
\draw (265.92,62.31) node   [align=left] {\begin{minipage}[lt]{30.26pt}\setlength\topsep{0pt}
\begin{center}
\textcolor[rgb]{1,1,1}{{\scriptsize sha()}}
\end{center}

\end{minipage}};

\end{tikzpicture}
    \end{center}
      \vspace{-0.1in}
      \caption{ECC Core Accelerator SoC.}
        \vspace{-0.2in}
      \label{fig:eccSoC}
\end{figure}
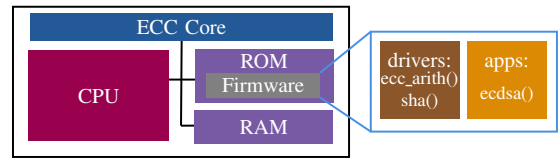

Elliptic Curve Cryptography (ECC) functions are popularly used in public key cryptosystems that can be utilized for both authentication and authenticated encryption. ECC core accelerator provides functionality to perform calculations related to the Elliptic Curve Digital Signature Algorithm (ECDSA) and Elliptic Curve Integrated Encryption Scheme (ECIES) faster.
Implementation of ECC core consists of  \textit{ECC\_Arithmetic core} and \textit{SHA256}. Users can use the ECC core as an IP in the SoC design as illustrated in Figure~\ref{fig:eccSoC}. Through the firmware driver, each of the functionalities can be accessed by the application. ECC core is connected with the SoC via the MMIO, and based on the request register status, it will perform the relevant calculation and the results will be written back to the results register. ECC core implementation consists of several nested FSMs in order to achieve the above functionalities. In order to evaluate the effectiveness of \textit{HIVE}, we have selected several functional scenarios from the \textit{ECC\_Arithmetic core}. These functional scenarios include testing the field addition, field multiplication, field inverse, and fast reduction. Table~\ref{tab:ECCresults} presents the elapsed time and consumed memory for hint generation as well as verification with \textit{HIVE}.

\begin{table}[htp]
\centering
\small
\caption{Time (in Minutes) and memory (in Megabytes) consumed by the verification process of the SoC with the ECC Core for different verification scenarios of \textit{Arithmetic Core}. \textit{We did not show comparison with state-of-the-art since existing methods without hints face timeouts for all scenarios.}}\label{tab:ECCresults}
\vspace{-0.05in}
\resizebox{\columnwidth}{!}{
\begin{tabular}{|l|cc|cc|}
\hline
\multicolumn{1}{|c|}{\multirow{2}{*}{\textbf{\begin{tabular}[c]{@{}c@{}}Test Cases for\\ Arithmetic Core\end{tabular}}}} & \multicolumn{2}{c|}{\textbf{Hint Generation}} & \multicolumn{2}{c|}{\textbf{Equival. Checking}} \\ \cline{2-5} 
\multicolumn{1}{|c|}{} & \multicolumn{1}{c|}{\textbf{Time}} & \multicolumn{1}{c|}{\textbf{Memory}} & \multicolumn{1}{c|}{\textbf{Time}} & \multicolumn{1}{c|}{\textbf{Memory}} \\ \hline
\textit{Field Addition} & \multicolumn{1}{c|}{33} &  2109& \multicolumn{1}{c|}{207} & 1876 \\ \hline
\textit{Field Multiplication} & \multicolumn{1}{c|}{47} & 2967 & \multicolumn{1}{c|}{239} & 1908 \\ \hline
\textit{Field Inverse} & \multicolumn{1}{c|}{41} & 2479 & \multicolumn{1}{c|}{211} & 1856 \\ \hline
\textit{Fast Reduction} & \multicolumn{1}{c|}{28} &  1795 & \multicolumn{1}{c|}{242} & 2064 \\ \hline
\end{tabular}}
\vspace{-0.1in}
\end{table}

\subsection{Case Study 3: LCD Controller}\label{LCDController}

\begin{figure}[htp]
    \begin{center}
        \tikzset{every picture/.style={line width=0.75pt}} 

\begin{tikzpicture}[x=0.75pt,y=0.75pt,yscale=-0.8,xscale=0.9]

\draw  [color={rgb, 255:red, 74; green, 144; blue, 226 }  ,draw opacity=1 ][line width=0.75]  (238.5,34.14) -- (342.83,34.14) -- (342.83,97.18) -- (238.5,97.18) -- cycle ;
\draw  [draw opacity=0][fill={rgb, 255:red, 93; green, 90; blue, 90 }  ,fill opacity=1 ] (238,3.02) -- (342.67,3.02) -- (342.67,27.48) -- (238,27.48) -- cycle ;

\draw    (132.17,77.45) -- (132.57,15.89) -- (140.86,15.89) ;
\draw    (120.66,48.35) -- (140.05,48.36) ;
\draw    (132.17,77.45) -- (141.5,77.4) ;
\draw  [draw opacity=0][fill={rgb, 255:red, 153; green, 0; blue, 77 }  ,fill opacity=1 ][line width=0.75]  (46.6,29.74) -- (125.5,29.74) -- (125.5,87.01) -- (46.6,87.01) -- cycle ;

\draw  [draw opacity=0][fill={rgb, 255:red, 120; green, 89; blue, 163 }  ,fill opacity=1 ][line width=0.75]  (139.85,29.51) -- (217.5,29.51) -- (217.5,63.02) -- (139.85,63.02) -- cycle ;
\draw  [draw opacity=0][fill={rgb, 255:red, 120; green, 89; blue, 163 }  ,fill opacity=1 ][line width=0.75]  (139.79,67.66) -- (217.5,67.66) -- (217.5,88.77) -- (139.79,88.77) -- cycle ;

\draw   (38,2.41) -- (226.33,2.41) -- (226.33,97.18) -- (38,97.18) -- cycle ;
\draw  [draw opacity=0][fill={rgb, 255:red, 128; green, 128; blue, 128 }  ,fill opacity=1 ][line width=0.75]  (146.43,44.11) -- (210.04,44.11) -- (210.04,59.62) -- (146.43,59.62) -- cycle ;

\draw  [draw opacity=0][fill={rgb, 255:red, 35; green, 89; blue, 150 }  ,fill opacity=1 ][line width=0.75]  (47.8,6.87) -- (217.11,6.87) -- (217.11,23.91) -- (47.8,23.91) -- cycle ;

\draw    (217.5,15.49) -- (237.78,15.49) ;
\draw  [draw opacity=0][fill={rgb, 255:red, 139; green, 87; blue, 42 }  ,fill opacity=1 ][line width=0.75]  (244,40.2) -- (288.5,40.2) -- (288.5,88.79) -- (244,88.79) -- cycle ;
\draw  [draw opacity=0][fill={rgb, 255:red, 220; green, 138; blue, 4 }  ,fill opacity=1 ][line width=0.75]  (292.5,40.2) -- (337,40.2) -- (337,88.8) -- (292.5,88.8) -- cycle ;

\draw [color={rgb, 255:red, 74; green, 144; blue, 226 }  ,draw opacity=1 ][line width=0.75]    (210.04,59.62) -- (237.5,97.18) ;
\draw [color={rgb, 255:red, 74; green, 144; blue, 226 }  ,draw opacity=1 ][line width=0.75]    (210.04,44.11) -- (238.22,34.14) ;

\draw (178.75,36.62) node   [align=left] {\begin{minipage}[lt]{52.7pt}\setlength\topsep{0pt}
\begin{center}
\textcolor[rgb]{1,1,1}{{\footnotesize ROM}}
\end{center}

\end{minipage}};
\draw (132.46,15.44) node   [align=left] {\begin{minipage}[lt]{115.13pt}\setlength\topsep{0pt}
\begin{center}
\textcolor[rgb]{1,1,1}{{\footnotesize LCD Controller}}
\end{center}

\end{minipage}};
\draw (178.23,51.91) node   [align=left] {\begin{minipage}[lt]{43.26pt}\setlength\topsep{0pt}
\begin{center}
\textcolor[rgb]{1,1,1}{{\footnotesize Firmware}}
\end{center}

\end{minipage}};
\draw (178.64,78.28) node   [align=left] {\begin{minipage}[lt]{52.85pt}\setlength\topsep{0pt}
\begin{center}
\textcolor[rgb]{1,1,1}{{\footnotesize RAM}}
\end{center}

\end{minipage}};
\draw (86.05,58.56) node   [align=left] {\begin{minipage}[lt]{53.65pt}\setlength\topsep{0pt}
\begin{center}
\textcolor[rgb]{1,1,1}{{\footnotesize CPU}}
\end{center}

\end{minipage}};
\draw (314.75,64.66) node   [align=left] {\begin{minipage}[lt]{30.26pt}\setlength\topsep{0pt}
\begin{center}
\textcolor[rgb]{1,1,1}{{\footnotesize apps:}}\\\textcolor[rgb]{1,1,1}{{\scriptsize graphics()}}
\end{center}

\end{minipage}};
\draw (290.16,15.98) node   [align=left] {\begin{minipage}[lt]{70.94pt}\setlength\topsep{0pt}
\begin{center}
\textcolor[rgb]{1,1,1}{{\footnotesize External Display}}
\end{center}

\end{minipage}};
\draw (266.25,64.6) node   [align=left] {\begin{minipage}[lt]{30.26pt}\setlength\topsep{0pt}
\begin{center}
\textcolor[rgb]{1,1,1}{{\footnotesize drivers:}}\\\textcolor[rgb]{1,1,1}{{\scriptsize display()}}
\end{center}

\end{minipage}};

\end{tikzpicture}
    \end{center}
      \vspace{-0.1in}
      \caption{LCD Controller SoC.}
        \vspace{-0.1in}
      \label{fig:lcdSoC}
\end{figure}
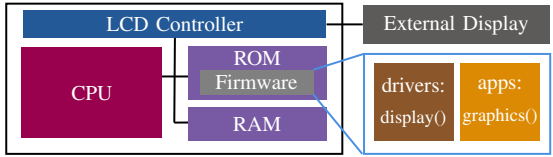

This SoC implementation consists of a hardware Liquid Crystal Display (LCD) controller as illustrated in Figure~\ref{fig:lcdSoC}. It can be configured by the firmware and then display data can be sent to the display. Configure parameters are sent as command packets from the firmware  and data packets are sent as data packets to the controller. This implementation is designed to work with ST77XX family display drivers. The controller consists of an FSM with 33 states and 30 transitions. ST77XX family device drivers have particular delay periods between different command parameters and data packets.

According to the specification, implementation requires satisfying several basic conditions to properly initialize the external display and to show graphics on the display. These requirements can be outlined as follows, 1) display delay values are within the specification range, 2) correct commands are sent to the external display from the firmware, and 3) data packets are in sync with the coordinates of the display. Based on the requirements, we formulated several test scenarios for this experiment. For each of the test scenarios, we have written one test case in the firmware. Then we used the original specification~\cite{st77xx} of the display driver to write the specification in \textit{Racket} language.

\begin{table}[htp]
\small
\caption{Time (in Minutes) and memory (in Megabytes) consumed by the verification process of the SoC with the SPI LCD Controller for different verification scenarios. \textit{We did not show comparison with state-of-the-art since existing methods without hints face timeouts for all scenarios.}}\label{tab:spiLCDresults}
\vspace{-0.05in}
\centering
\resizebox{\columnwidth}{!}{
\begin{tabular}{|l|l|cc|cc|}
\hline
\multicolumn{1}{|c|}{\multirow{2}{*}{\textbf{Property}}} & \multicolumn{1}{c|}{\multirow{2}{*}{\textbf{Test Cases}}} & \multicolumn{2}{c|}{\textbf{\begin{tabular}[c]{@{}c@{}}Hint\\ Generation\end{tabular}}} & \multicolumn{2}{c|}{\textbf{\begin{tabular}[c]{@{}c@{}}Equivalence\\ Checking\end{tabular}}} \\ \cline{3-6} 
\multicolumn{1}{|c|}{} & \multicolumn{1}{c|}{} & \multicolumn{1}{c|}{\textbf{Time}} & \multicolumn{1}{c|}{\textbf{Mem}} & \multicolumn{1}{c|}{\textbf{Time}} & \multicolumn{1}{c|}{\textbf{Mem}} \\ \hline
\multirow{4}{*}{\begin{tabular}[c]{@{}l@{}}Delay\\ Check\end{tabular}} & \textit{SWRESET} & \multicolumn{1}{c|}{23} & 1649 & \multicolumn{1}{c|}{37} &  1489 \\\cline{2-6} 
 & \textit{SLPOUT} & \multicolumn{1}{c|}{29} & 1720 & \multicolumn{1}{c|}{33} & 1348 \\ \cline{2-6} 
 & \textit{NORON} & \multicolumn{1}{c|}{27} & 1644 & \multicolumn{1}{c|}{32} & 1394 \\ \cline{2-6} 
 & \textit{DISPON} & \multicolumn{1}{c|}{27} & 1679 & \multicolumn{1}{c|}{42} & 1489 \\ \hline
\multirow{4}{*}{\begin{tabular}[c]{@{}l@{}}Command\\ Mode\end{tabular}} & \textit{CASET} & \multicolumn{1}{c|}{19} & 976 & \multicolumn{1}{c|}{23} & 1296 \\ \cline{2-6} 
 & \textit{RASET} & \multicolumn{1}{c|}{19} & 981 & \multicolumn{1}{c|}{22} & 1301 \\ \cline{2-6} 
 & \textit{COLMOD} & \multicolumn{1}{c|}{19} & 967 & \multicolumn{1}{c|}{21} & 1290 \\ \cline{2-6} 
 & \textit{MADCTL} & \multicolumn{1}{c|}{20} &  970& \multicolumn{1}{c|}{26} & 1310 \\ \hline
\multirow{3}{*}{\begin{tabular}[c]{@{}l@{}}Data\\ Mode\end{tabular}} & \textit{NEW\_FRAME} & \multicolumn{1}{c|}{17} & 979 & \multicolumn{1}{c|}{31} & 1358 \\ \cline{2-6} 
 & \textit{WAIT\_FRAME} & \multicolumn{1}{c|}{19} & 963 & \multicolumn{1}{c|}{33} & 1380 \\ \cline{2-6} 
 & \textit{INIT\_FRAME} & \multicolumn{1}{c|}{18} & 958 & \multicolumn{1}{c|}{33} & 1367 \\ \hline
\end{tabular}}
\end{table}

\begin{figure}[htp]
  
    \begin{subfigure}{1\linewidth}
      \centering   \includegraphics[width=1\columnwidth]{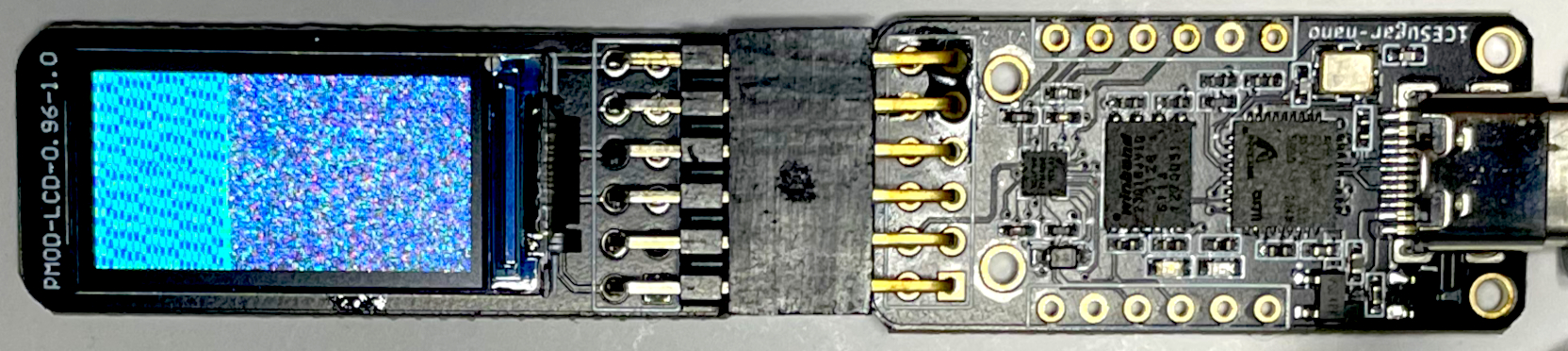}
      \vspace{-0.2in}
      \caption{SoC with the bug in driver code, before the verification process.}
      \label{fig:displaybug}

    \end{subfigure}%

    \begin{subfigure}{1\linewidth}
      \centering
    \includegraphics[width=1\columnwidth]{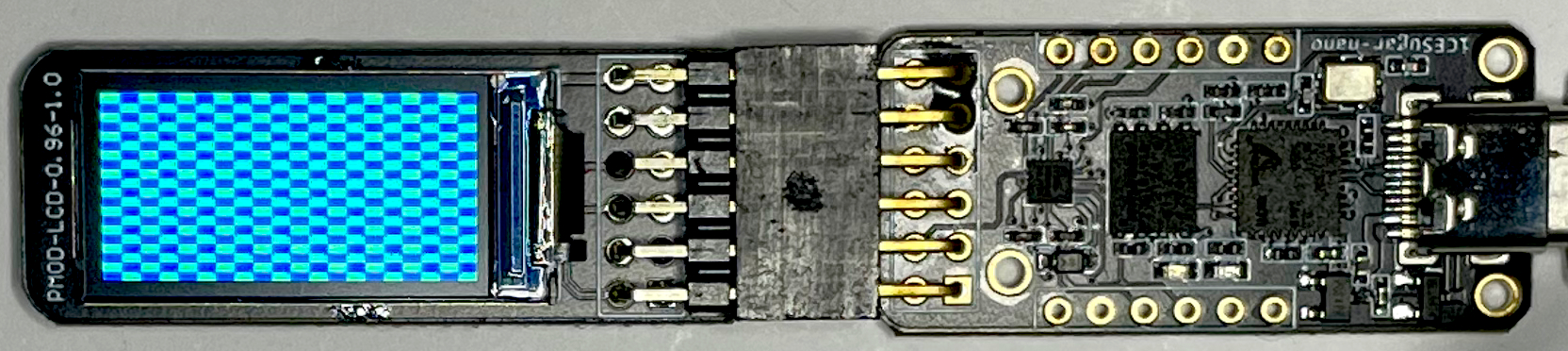}
      \vspace{-0.2in}
      \caption{SoC after the verification process and fixing the driver bug.}
      \label{fig:bugfixed}

    \end{subfigure}%

      \caption{SoC with SPI LCD controller running \textit{checkered\_Flag.c} benchmark. Firmware is sending the display content to the  external display with the ST7735 driver on \textit{Lattice Ice-Sugar-Nano} FPGA.}
      \vspace{-0.15in}
      \label{fig:st7735}
\end{figure}

%
After applying the \textit{HIVE} on the test scenarios, we performed equivalence checking to provide a guarantee for all the scenarios. Table~\ref{tab:spiLCDresults} illustrates the verification scenarios for the LCD controller SoC and elapsed time and memory used for each of the verification sub-problem with \textit{HIVE} generated hints. We were able to reveal a functional bug related to \textit{CASET} and \textit{RASET} test cases that were causing the external ST7735 driver to set up X and Y coordinate values incorrectly, this was fixed in the firmware device driver. Figure~\ref{fig:st7735} illustrates the design running on the FPGA with  \textit{checkered\_Flag.c} benchmark. Figure~\ref{fig:displaybug} illustrates the SoC with the driver bug, and here the external display displays graphics incorrectly. Figure~\ref{fig:bugfixed} presents the validated SoC after fixing the driver bug, displaying data on the external display properly.

\vspace{-0.1in}
\subsection{Case Study 4: Mass Storage (SATA) Controller}\label{subsec:sata}

\begin{figure}[htp]
    \begin{center}
    \vspace{-0.1in}
        \tikzset{every picture/.style={line width=0.75pt}} 

\begin{tikzpicture}[x=0.75pt,y=0.75pt,yscale=-0.8,xscale=0.9]

\draw  [color={rgb, 255:red, 74; green, 144; blue, 226 }  ,draw opacity=1 ][line width=0.75]  (238.5,34.14) -- (342.83,34.14) -- (342.83,97.18) -- (238.5,97.18) -- cycle ;
\draw  [draw opacity=0][fill={rgb, 255:red, 93; green, 90; blue, 90 }  ,fill opacity=1 ] (238,3.02) -- (342.67,3.02) -- (342.67,27.48) -- (238,27.48) -- cycle ;

\draw    (132.17,77.45) -- (132.57,15.89) -- (140.86,15.89) ;
\draw    (120.66,48.35) -- (140.05,48.36) ;
\draw    (132.17,77.45) -- (141.5,77.4) ;
\draw  [draw opacity=0][fill={rgb, 255:red, 153; green, 0; blue, 77 }  ,fill opacity=1 ][line width=0.75]  (46.6,29.74) -- (125.5,29.74) -- (125.5,87.01) -- (46.6,87.01) -- cycle ;

\draw  [draw opacity=0][fill={rgb, 255:red, 120; green, 89; blue, 163 }  ,fill opacity=1 ][line width=0.75]  (139.85,29.51) -- (217.5,29.51) -- (217.5,63.02) -- (139.85,63.02) -- cycle ;
\draw  [draw opacity=0][fill={rgb, 255:red, 120; green, 89; blue, 163 }  ,fill opacity=1 ][line width=0.75]  (139.79,67.66) -- (217.5,67.66) -- (217.5,88.77) -- (139.79,88.77) -- cycle ;

\draw   (38,2.41) -- (226.33,2.41) -- (226.33,97.18) -- (38,97.18) -- cycle ;
\draw  [draw opacity=0][fill={rgb, 255:red, 128; green, 128; blue, 128 }  ,fill opacity=1 ][line width=0.75]  (146.43,44.11) -- (210.04,44.11) -- (210.04,59.62) -- (146.43,59.62) -- cycle ;

\draw  [draw opacity=0][fill={rgb, 255:red, 35; green, 89; blue, 150 }  ,fill opacity=1 ][line width=0.75]  (47.8,6.87) -- (217.11,6.87) -- (217.11,23.91) -- (47.8,23.91) -- cycle ;

\draw    (217.5,15.49) -- (237.78,15.49) ;
\draw  [draw opacity=0][fill={rgb, 255:red, 139; green, 87; blue, 42 }  ,fill opacity=1 ][line width=0.75]  (244,40.2) -- (288.5,40.2) -- (288.5,88.79) -- (244,88.79) -- cycle ;
\draw  [draw opacity=0][fill={rgb, 255:red, 220; green, 138; blue, 4 }  ,fill opacity=1 ][line width=0.75]  (292.5,40.2) -- (337,40.2) -- (337,88.8) -- (292.5,88.8) -- cycle ;

\draw [color={rgb, 255:red, 74; green, 144; blue, 226 }  ,draw opacity=1 ][line width=0.75]    (210.04,59.62) -- (237.5,97.18) ;
\draw [color={rgb, 255:red, 74; green, 144; blue, 226 }  ,draw opacity=1 ][line width=0.75]    (210.04,44.11) -- (238.22,34.14) ;

\draw (178.75,36.62) node   [align=left] {\begin{minipage}[lt]{52.7pt}\setlength\topsep{0pt}
\begin{center}
\textcolor[rgb]{1,1,1}{{\footnotesize ROM}}
\end{center}

\end{minipage}};
\draw (132.46,15.44) node   [align=left] {\begin{minipage}[lt]{115.14pt}\setlength\topsep{0pt}
\begin{center}
\textcolor[rgb]{1,1,1}{{\footnotesize SATA Controller}}
\end{center}

\end{minipage}};
\draw (178.23,51.91) node   [align=left] {\begin{minipage}[lt]{43.26pt}\setlength\topsep{0pt}
\begin{center}
\textcolor[rgb]{1,1,1}{{\footnotesize Firmware}}
\end{center}

\end{minipage}};
\draw (178.64,78.28) node   [align=left] {\begin{minipage}[lt]{52.85pt}\setlength\topsep{0pt}
\begin{center}
\textcolor[rgb]{1,1,1}{{\footnotesize RAM}}
\end{center}

\end{minipage}};
\draw (86.05,58.56) node   [align=left] {\begin{minipage}[lt]{53.65pt}\setlength\topsep{0pt}
\begin{center}
\textcolor[rgb]{1,1,1}{{\footnotesize CPU}}
\end{center}

\end{minipage}};
\draw (314.75,64.66) node   [align=left] {\begin{minipage}[lt]{30.26pt}\setlength\topsep{0pt}
\begin{center}
\textcolor[rgb]{1,1,1}{{\footnotesize apps:}}\\\textcolor[rgb]{1,1,1}{{\scriptsize media()}}
\end{center}

\end{minipage}};
\draw (290.16,15.98) node   [align=left] {\begin{minipage}[lt]{70.94pt}\setlength\topsep{0pt}
\begin{center}
\textcolor[rgb]{1,1,1}{{\footnotesize Mass Storage}}
\end{center}

\end{minipage}};
\draw (266.25,64.6) node   [align=left] {\begin{minipage}[lt]{30.26pt}\setlength\topsep{0pt}
\begin{center}
\textcolor[rgb]{1,1,1}{{\footnotesize drivers:}}\\\textcolor[rgb]{1,1,1}{{\scriptsize sata()}}
\end{center}

\end{minipage}};

\end{tikzpicture}
    \end{center}
      \vspace{-0.1in}
      \caption{SoC that integrates SATA controller to communicate with the mass storage device.}
        \vspace{-0.1in}
      \label{fig:sataSoC}
\end{figure}
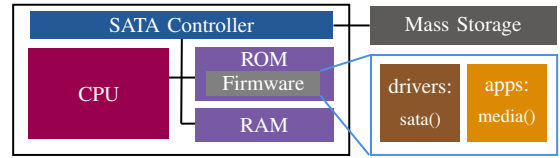

Serial Advanced Technology Attachment (SATA) controller is a hardware module that manages the communication between a SoC and mass storage devices (such as hard disk drives and solid-state drives). In this case study, we have utilized a SATA controller that connects to the SoC using the AXI (Advanced eXtensible Interface) as illustrated in Figure~\ref{fig:sataSoC}. The controller is configured and managed by the firmware driver, which has the ability to change the state of the data transfer between the mass storage device attached to the controller based on the requirement of the application code. For simulation purposes, we have used the Verilog \texttt{readmemb}  function as the mass storage device.

The firmware driver is responsible for managing the SATA controller to perform handshake, establishing and maintaining the communication protocol between the system and the storage device, thereby ensuring reliable and synchronized data exchange. SATA controller implementation consists of several nested FSMs in order to achieve the above functionalities. In order to evaluate the effectiveness of \textit{HIVE}, we have selected several functional scenarios from the implementation. These functional scenarios include testing the controller initialization, SATA handshake, controller commands, and read and write transactions. Table~\ref{tab:SATAresults} presents the elapsed time and consumed memory for hint generation as well as verification with \textit{HIVE}.

\begin{table}[t]
\centering
\small
\caption{Time (in Minutes) and memory (in Megabytes) consumed by the verification process of the SoC with the SATA device connected via SATA controller for different verification scenarios of \textit{SATA controller}. \textit{We did not show comparison with state-of-the-art since existing methods without hints face timeouts for all scenarios.}}\label{tab:SATAresults}
\vspace{-0.05in}
\resizebox{\columnwidth}{!}{
\begin{tabular}{|l|cc|cc|}
\hline
\multicolumn{1}{|c|}{\multirow{2}{*}{\textbf{\begin{tabular}[c]{@{}c@{}}Test Cases for\\ SATA Driver\end{tabular}}}} & \multicolumn{2}{c|}{\textbf{Hint Generation}} & \multicolumn{2}{c|}{\textbf{Equival. Checking}} \\ \cline{2-5} 
\multicolumn{1}{|c|}{} & \multicolumn{1}{c|}{\textbf{Time}} & \multicolumn{1}{c|}{\textbf{Memory}} & \multicolumn{1}{c|}{\textbf{Time}} & \multicolumn{1}{c|}{\textbf{Memory}} \\ \hline
\textit{Initialization Sequence} & \multicolumn{1}{c|}{39} &  3109& \multicolumn{1}{c|}{274} & 2850 \\ \hline
\textit{Handshake Phase} & \multicolumn{1}{c|}{41} & 3190 & \multicolumn{1}{c|}{269} & 2689 \\ \hline
\textit{Command Phase} & \multicolumn{1}{c|}{43} & 3019 & \multicolumn{1}{c|}{255} & 2503 \\ \hline
\textit{RX Transaction} & \multicolumn{1}{c|}{34} &  3100 & \multicolumn{1}{c|}{270} & 2476 \\ \hline
\textit{TX Transaction} & \multicolumn{1}{c|}{34} &  3100 & \multicolumn{1}{c|}{270} & 2476 \\ \hline
\end{tabular}}
\vspace{-0.1in}
\end{table}

\color{black}

\section{Applicability and Limitations}\label{sec:limits}
In this paper, we focused on automated generation of hints to simplify equivalence checking during functional verification by reducing the state space. In this section, we discuss the applicability and limitations of \textit{HIVE}.

\vspace{0.1in}
\noindent \underline{Usability and Tools}: \textit{HIVE} usability is not limited to functional verification. This framework can be used in other scenarios (e.g., security validation) to reduce the state space and make the problem more tractable. Note that Rosette was used in the experiments due to seamless integration with the backend SMT (Z3) solver for symbolic execution to aid formal proofs. However, other formal tools also can be used with minor modifications to the framework. 

\vspace{0.1in}
\noindent \underline{Simulation versus Formal Guarantee}:
Simulation does not provide any verification guarantee since it is infeasible to simulate using all possible input sequences. The contribution of \textit{HIVE} is to translate one concrete test case to a symbolic test case and perform equivalence checking to obtain formal guarantee. For example, consider the addition functionality (scenario) of a 64-bit calculator. In order to verify addition, \textit{HIVE} only needs one test case, such as $2 + 3 = 5$. By looking at the simulation trace, Algorithm~\ref{algo:hintsynth} figures out which paths are relevant for the hint generation, which then facilitates symbolic simulation of the addition scenario.  During the symbolic simulation, symbolic values will propagate through the design covering all possible inputs (instead of concrete values like $2+5$). This provides guarantees for addition functionality for all the inputs of $2^{64} \times 2^{64}$. If our framework is not used, a designer has to simulate the addition scenario of the calculator $2^{64} \times 2^{64}$ times to get the same level of verification guarantee.

\vspace{0.1in}
\noindent \underline{Effort for Implementation Abstraction}: The manual specification requirement remains the same for both \textit{HIVE} and existing (state-of-the-art) approaches. The fundamental difference is that existing approach requires manual abstraction of the implementation. For example, ``C=A+B'' would be the specification of an adder.  The abstraction of the adder implementation will look different depending on whether it was implemented as ripple-carry adder or carry look-ahead adder. In contrast, our framework eliminates the need for manually writing a per-scenario abstraction of the implementation. In other words, the automatically generated hints slices the implementation to its smallest possible size during symbolic simulation, giving the same effect as manually abstracted implementation.

\begin{figure}[H]
    \begin{center}
    \vspace{-0.1in}
        \input{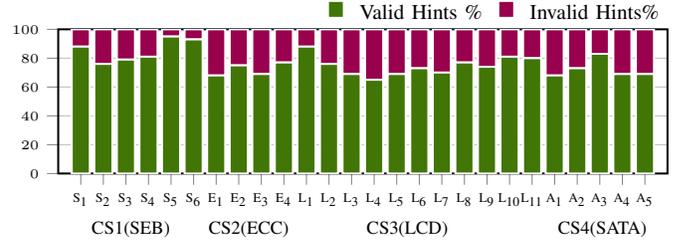}
    \end{center}
      \vspace{-0.1in}
      \caption{Percentage of valid and invalid hints generated by \textit{HIVE} framework for different scenarios of four case studies. Here S\textsubscript{1} to S\textsubscript{6}, E\textsubscript{1} to E\textsubscript{4}, L\textsubscript{1} to L\textsubscript{11} and  A\textsubscript{1} to A\textsubscript{5} denotes scenarios corresponding case studies CS1, CS2, CS3, and CS4.}
      \vspace{-0.2in}
      \label{fig:validinvalid}
\end{figure}

\vspace{0.1in}
\noindent \underline{Invalid Hints}:
Since \textit{HIVE} is focused on automating the manual hint generation procedure, it also has some of the inherent drawbacks of the manual method, such as the generation of invalid hints. However, the percentage of invalid hints is typically small. Note that if hints are not available at all, SMT will face state space explosion. Figure~\ref{fig:validinvalid} shows the percentage of valid and invalid hints generated by  HIVE for each of the validation scenarios. It can be observed that the scenarios evaluated in the {four} case studies (CS1:S1-S6, CS2:E1-E4,  CS3:L1-L11, and {CS4:A1-A5)}, the vast majority of the hints generated by \textit{HIVE} are valid hints. Note that the proof is still complete since there will be no false positives, and the false negatives can be eliminated by utilizing the counterexamples.

\vspace{0.1in}
\noindent \underline{False-Negative Results}: \textit{HIVE} does not produce false-positive results since the hardware implementation is directly converted to an SMT model and it has the exact SMT model of the implementation. Both \textit{HIVE} and state-of-the-art manual abstraction methods may produce false negatives if the specification is incorrect or incomplete. The false-negative results will not directly affect the verification guarantees of the implementation since the symbolic evaluation results will provide a counterexample. A designer can analyze the counterexample to find it as a false negative and can modify the specification to fix the problem. 

\section{Conclusion} 
\label{sec:conclusion}

Hardware-firmware co-verification is an important requirement in designing reliable systems. While there are promising formal verification approaches, they rely on manually written hints or abstracted versions of the hardware to deal with the state space explosion. Manual intervention requires design expertise and can be time-consuming and error-prone.
In this paper, we presented an automated hint-based verification (\textit{HIVE}) framework that does not require any manual intervention during model generation or hint extraction. We effectively utilize both static analysis and concrete simulation of the implementation to extract proof supporting artifacts from the system model. \textit{HIVE} can automatically generate system-level hints to guide formal proofs to avoid state space explosion. Extensive evaluation using real-world hardware-firmware implementations demonstrates the effectiveness and scalability of the proposed framework.  In all the case studies, \textit{HIVE} was able to provide a guarantee for the scenarios in a reasonable time, while state-of-the-art methods failed to verify them (timeout). Moreover, \textit{HIVE} is able to identify several faulty scenarios in the hardware-firmware implementations.


\bibliographystyle{IEEEtran}
\bibliography{IEEEabrv,bibliography}

\vspace{-0.5in}
\begin{IEEEbiography}[{\includegraphics[width=1in,clip]{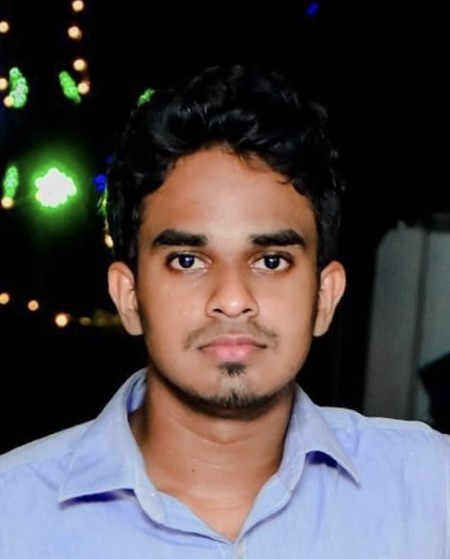}}]{Aruna Jayasena} is a Ph.D student in the Department of Computer \& Information Science \& Engineering at the University of Florida. He received his B.S. in the Department of Computer Science and Engineering at the University of Moratuwa, Sri Lanka, in 2019. His research focuses on systems security, hardware-firmware co-validation, test generation, trusted execution, side-channel analysis, and system-on-chip debug.
\end{IEEEbiography}

\vspace{-0.5in}
\begin{IEEEbiography}[{\includegraphics[width=1in,clip,keepaspectratio]{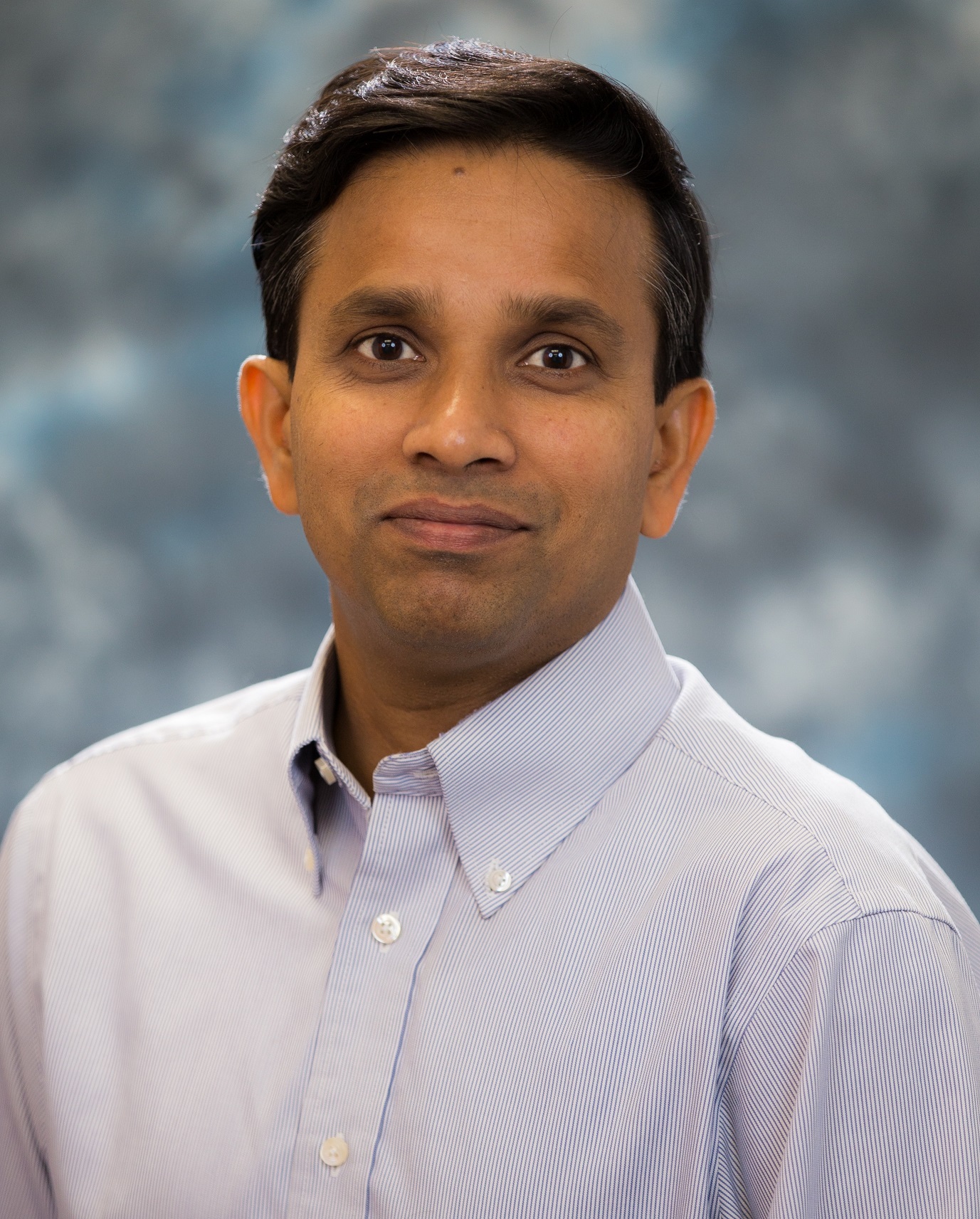}}]{Prabhat Mishra} is a Professor in the Department of Computer and Information Science and Engineering at the University of Florida. He received his Ph.D. in Computer Science from the University of California at Irvine. His research interests include embedded systems, hardware security, formal verification, system-on-chip validation, machine learning, and quantum computing. He currently serves as an Associate Editor of ACM Transactions on Embedded Computing Systems. He is an IEEE Fellow, an AAAS Fellow, and an ACM Distinguished Scientist.
\end{IEEEbiography}

\end{document}